\documentclass[review,3p, 12pt]{elsarticle}

\usepackage[version=4]{mhchem}
\usepackage{amsmath,amsmath,mathrsfs}
\usepackage{color,fancyhdr,epsf,psfrag,lineno,txfonts,subfigure}
\usepackage{multirow}
\usepackage{graphicx}
\usepackage{mhchem}
\graphicspath{ {figs/} }
\usepackage{bm,ulem}

\definecolor{R1}{rgb}{0.0,0.0,0.0} 
\definecolor{R3}{rgb}{0.0,0.0,0.0} 
\definecolor{R2}{rgb}{0.0,0,0.0} 

\usepackage{siunitx}
\usepackage{amssymb}

\journal{Elsevier}

\biboptions{sort&compress}

\begin{document}

\begin{frontmatter}

\title{Large-eddy simulation of turbulent spray flames: Effects of scalar correlation and enthalpy reduction in flamelet modeling}

\author[a]{Dong Wang}
\author[b]{Min Zhang}
\author[c,d]{Yan Zhang}
\author[a,e]{Ruixin Yang}
\author[a,e]{Zhi X. Chen\corref{cor1}}
	\cortext[cor1]{Corresponding author.}
	\ead{chenzhi@pku.edu.cn}

\address[a]{State Key Laboratory of Turbulence and Complex Systems, College of Engineering, Peking University, Beijing 100871, P.R. China}
\address[b]{School of Energy Science and Engineering, Central South University, Changsha, 410083, P.R. China}
\address[c]{CAEP Software Center for High Performance Numerical Simulation, Beijing 100088, P.R. China}
\address[d]{Institute of Applied Physics and Computational Mathematics, Beijing, 100088, P.R. China}
\address[e]{AI for Science Institute (AISI), Beijing, 100080, P.R. China}

\begin{abstract}
Numerical modeling of turbulent spray combustion provides a promising tool for advanced engine design. In spray flames, the droplet evaporation not only reduces the ambient gas temperature, but also influences flame structure by generating substantial local fluctuations of the mixture fraction $\widetilde{Z}$ and progress variable $\widetilde{c}$. These two scalars, conventionally assumed independent in flamelet models, exhibit significant correlations arising from the coupling among evaporation, turbulent mixing and chemical reactions. This study proposes a six-dimensional flamelet-generated manifolds (FGM) method, considering the evaporation-induced specific enthalpy reduction and scalar correlation. A novel joint presumed probability density function (PDF) method is derived using the copula theory, achieving rapid grid convergence and good feasibility. Large-eddy simulation (LES) is performed on the Sydney ethanol turbulent spray flames (EtF1, EtF4 and EtF7), which feature different ethanol mass flow rates and jet Reynolds numbers. Both gas and liquid phase statistics show good agreement with experimental data across the three flames. The incorporation of specific enthalpy reduction and scalar correlation in FGM modeling improves gas temperature predictions, along with enhanced liquid-phase prediction through refined gas-field resolution. The correlation coefficient of $\widetilde{Z}$ and $\widetilde{c}$ is found to be a competing result of local evaporation and combustion, since evaporation elevates $\widetilde{Z}$ and dilutes reaction products, whereas chemical reactions enhance $\widetilde{c}$ fluctuations.

\end{abstract}

\vspace{12 pt}
\begin{keyword}
Large-eddy simulation; Turbulent flame; Spray combustion; Joint PDF method
\end{keyword}

\end{frontmatter}


\section{Introduction}\label{introduction}

Turbulent spray combustion is widely applied in aerospace propulsion, rocket engines, and internal combustion systems, because liquid fuels exhibit advantages in high volumetric energy density, safe storage/transport, and operational flexibility~\cite{TENG2023113062, yin2024experimental}. 
In pursuit of sustainable and green aviation, ethanol is a promising alternative fuel for aero-engines due to its renewable and sustainable production, low freezing point, and reduced pollutant emissions~\cite{yin2024experimental, MENDIBURU2022124688}. 
Numerical simulation is a powerful tool to investigate flame structures and optimize engine design.

To study the effects of liquid fuel and air carrier mass flow rates on the flame structure, turbulent ethanol spray flames EtF1, EtF4 and EtF7 were experimentally investigated as part of the Sydney dilute spray flame series~\cite{GOUNDER20123372, Gounder2009PhD, masri2010turbulent}. The availability of experiment data and previously published simulation studies~\cite{de2013large, hu2019partially, huspray2017, hu2020large, rittler2015sydney, heye2013probability, el2016large, kirchmann2021two, hussien2022simulations, sacomano2020interaction, yi2022large, chrigui2013large} provides a solid foundation for the present study. Hussien et al.~\cite{hussien2022simulations} conducted simulations on EtF1, EtF3 and EtF4, a series of flames characterized by decreasing ethanol mass flow rates while maintaining a fixed air carrier velocity. They reported a significant underprediction of temperature in EtF1 and EtF3 upstream along the central axis. Possible sources of errors were identified: ($\romannumeral1$) the assumption of statistical independence in the probability density function (PDF) method and ($\romannumeral2$) the neglect of evaporation-induced specific enthalpy reduction $\widetilde{h}_{r}$ in the lookup table. To address these limitations, the present work aims to evaluate the impact of specific enthalpy reduction and scalar correlation on flame behavior.

Spray combustion is characterized by a wide range of length and time scales, as well as the intense coupling of evaporation, turbulence, and combustion across atomization, droplet transport, vaporization, and reaction processes~\cite{de2013large, hu2019partially}. To capture the instantaneous turbulence using affordable computational resources, large-eddy simulation (LES) has been widely applied, which resolves the large-scale turbulent structures and models those in small spatial scales, i.e., sub-grid scales (SGS)~\cite{ZHIYIN201511}. Another challenge in spray combustion modeling lies in the multi-composition of liquid fuels. Compared to gaseous fuels like hydrogen, ammonia, and methane, liquid fuels involve more carbon and hydrogen atoms, leading to intricate chemical reaction mechanisms comprising hundreds of species and thousands of reactions~\cite{LAW20071}. Resolving detailed mechanisms poses a significant computational cost, particularly for industrial applications. To overcome this issue, various combustion models have been developed to project the high-dimensional thermochemical state space to a low-dimensional manifold while preserving the essential reaction dynamics~\cite{ihme2012regu}. Among them, the flamelet-generated manifolds (FGM) combustion model has been successfully applied to several spray combustion configurations~\cite{sulanumerical2023, sula2022large, lucchini2020modeling}, demonstrating both high accuracy and computational efficiency. 

A key advantage of the FGM approach is that it enables the reconstruction of thermo-chemical states with only a limited set of transported scalars, significantly reducing computational costs through table lookup. Consequently, the selection of transported scalars and their corresponding SGS closure is central to the accuracy and efficiency of the model. As the flow evolves, the Favre-averaged mixture fraction $\widetilde{Z}$ is a tracking scalar for the large-scale mixing of fuel, oxidizer and products. To resolve the chemical reaction states, the reaction progress variable $\widetilde{c}$ is introduced. The specific enthalpy $\widetilde{h}$ is transported to account for two-phase energy exchange. Further, refinements in turbulent-chemistry interaction (TCI) modeling require the transport of the mixture fraction variance, $\widetilde{Z^{''2}}$, and progress variable variance, $\widetilde{c^{''2}}$, to capture SGS mixing and reaction fluctuation, respectively. In LES, the covariance $\widetilde{Z^{''} c^{''}}$ signifies the SGS correlation of $\widetilde{Z}$ and $\widetilde{c}$ fluctuations. Influences of $\widetilde{Z^{''} c^{''}}$ on the flame thickness and reaction rate have been revealed in the stratified or partially premixed piloted jet flames via experiments~\cite{barlow2017defining, robin2008experimental} and direct numerical simulations (DNS)~\cite{jaganath2021transported, chen2018priori}. However, the \textit{posteriori} modeling of covariance remains challenging. The correlation effect needs to be included in the PDF method to construct the lookup table. The PDF method reported in Ref.~\cite{ruan2014modelling} requires over 5000 sampling points in $\widetilde{Z}$ and $\widetilde{c}$ space respectively, which is costly to create a high-dimensional table. To overcome this issue, this work proposes a more efficient joint PDF method that improves computational feasibility while maintaining accuracy in covariance modeling. 


Considering the above challenges, the main objectives of this paper are: ($\romannumeral1$) to formulate a joint PDF method for the SGS correlation of $\widetilde{Z}$ and $\widetilde{c}$ fluctuations; and ($\romannumeral2$) to illustrate the effects of evaporation-induced specific enthalpy reduction and scalar correlation on the Sydney ethanol spray flames~\cite{GOUNDER20123372, Gounder2009PhD}, using the high-dimensional FGM model in LES. To the best of our knowledge, this is the first study to employ 6-dimensional FGM modeling for the LES of turbulent spray flames, where the various subgrid processes of mixing, reaction, specific enthalpy reduction and scalar correlation, and their interactions can be systematically examined.
The remainder of this paper is organized as follows. The modeling approaches will be described in Section 2. Computational setups will be explained in Section 3. Results will be presented in Section 4. The conclusion will be given in Section 5.

\section{Methodology}\label{Methodology}

For the two-phase simulation, an Eulerian-Lagrangian framework is adopted in this study. The LES turbulence model and FGM model are used to resolve the instantaneous evolution of the flame structure. The liquid phase is modeled as Lagrangian particles which consist of droplet parcels.

\subsection{LES governing equations for continuous phase}\label{LES_governing_eqn}

The LES-filtered governing equations of mass, momentum and enthalpy are
\begin{equation}\label{mass_eqn}
\begin{aligned}
&\frac{\partial \overline{\rho}}{\partial t} + \frac{\partial \overline{\rho} \widetilde{u}_j }{\partial x_j} = \overline{\dot{S}}_{v},
\end{aligned}
\end{equation}

\begin{equation}\label{momentum_eqn}
\begin{aligned}
&\frac{\partial \overline{\rho} \widetilde{u}_i}{\partial t} + \frac{\partial \overline{\rho} \widetilde{u}_i \widetilde{u}_j}{\partial x_j} = - \frac{\partial \overline{p}}{\partial x_i} + \frac{\partial \left( \overline{\tau}_{ij} + \overline{\tau}_{sgs} \right)}{\partial x_j} + \overline{\rho} g_i + \overline{\dot{S}}_{m,i},
\end{aligned}
\end{equation}

\begin{equation}\label{enthalpy_eqn}
\begin{aligned}
&\frac{\partial \overline{\rho} \widetilde{h}}{\partial t} + \frac{\partial \overline{\rho} \widetilde{u}_j \widetilde{h}}{\partial x_j} = \frac{\partial}{\partial x_j} \left[ \overline{\rho} \left( \widetilde{\alpha}_h + \widetilde{\alpha}_{h,t} \right) \frac{\partial \widetilde{h}}{\partial x_j} \right] + \frac{D \overline{p}}{D t} + \overline{\dot{S}}_{h},
\end{aligned}
\end{equation}
where $\rho$ is the density, $t$ denotes the time, $u_j$ represents the velocity component in spatial direction $x_j$ ($j=1,2,3$), $p$ is the absolute pressure, $g$ is the acceleration of gravity, $D$ stands for the material derivative, $\overline{(\cdot)}$ denotes the Reynolds average operation, and $\widetilde{(\cdot)}$ indicates the Favre average operation. The viscous tensor $\tau$ is filtered to the resolved part, $\overline{\tau}_{ij}$, and the unresolved SGS part, $\overline{\tau}_{sgs}$. In this study, the Sigma model~\cite{nicoud2011using} is applied to account for $\overline{\tau}_{sgs}$, because of the good prediction in turbulent viscosity for thermal expansion~\cite{rittler2015sydney}. The source terms $\overline{\dot{S}_v}$, $\overline{\dot{S}}_{m,i}$ and $\overline{\dot{S}}_{h}$ signify the two-phase mass, momentum and energy exchange, respectively, as given in Section~\ref{liquid_phase_model}. In Eq.~(\ref{enthalpy_eqn}), it is worth mentioning that the effects of radiation and viscous dissipation are neglected, and that the unity Lewis number assumption and Fick's Law are applied. The diffusivity $\widetilde{\alpha}_h = \widetilde{\kappa} / \widetilde{C}_p$, and $\widetilde{\alpha}_{h,t} = \overline{\rho} \widetilde{\nu}_t / Pr_t$, where $\widetilde{\kappa}$ is the thermal conductivity, $\widetilde{C}_p$ is the specific heat capacity, $\widetilde{\nu_t}$ is the turbulent kinematic viscosity, and $Pr_t$ is the turbulent Prandtl number. The gas-phase temperature is computed as
\begin{equation}\label{T_expr}
\begin{aligned}
&\widetilde{T} = T_0 + \left( \widetilde{h} - \Delta \widetilde{h}_{f}^o \right) / \widetilde{C}_{p,\text{eff}},
\end{aligned}
\end{equation}
where $T_0$ = 298K, and $\Delta \widetilde{h}_{f}^o$ is the Favre-filtered formation enthalpy. The effective specific heat capacity $C_{p,\text{eff}} = \left( \int_{T_0}^{T_1} C_p dT \right) \Big/ \left( T_1 - T_0 \right)$, where $T_1$ is the local temperature at which $C_{p,\text{eff}}$ is calculated. The local density is updated using the gas state equation, $\overline{\rho}=\overline{p} \widetilde{W}_{mix} / (R_0 \widetilde{T})$, where $\widetilde{W}_{mix}$ is the molecular weight of the mixture, and $R_0$ is the universal gas constant.

In this work, the FGM model is applied to characterize the thermochemical state in the two-phase flow. The mixture fraction $\widetilde{Z}$ is defined using Bilger's definition~\cite{bilger1990reduced}. The progress variable is given as a linear combination of four reaction products, namely, 
\begin{equation}\label{c_represent}
\begin{aligned}
&c = Y_c = Y_{H_2} + Y_{H_2 O} + Y_{CO} + Y_{CO_2},
\end{aligned}
\end{equation}
which is consistent with the previous LES studies~\cite{de2013large, rittler2015sydney, hu2018nonpremixed} on the Sydney ethanol spray flames. With the unity Lewis number assumption, the transport equations of the combustion scalars are written as
\begin{equation}\label{z_eqn}
\begin{aligned}
&\frac{\partial \overline{\rho} \widetilde{Z}}{\partial t} + \frac{\partial \overline{\rho} \widetilde{u}_j \widetilde{Z}}{\partial x_j} = \frac{\partial}{\partial x_j} \left[ \left( \overline{\rho} \widetilde{\alpha}_m+ \frac{\widetilde{\mu}_t}{Sc_t} \right) \frac{\partial \widetilde{Z}}{\partial x_j} \right] + \overline{\dot{S}_v},
\end{aligned}
\end{equation}

\begin{equation}\label{zvar_eqn}
\begin{aligned}
&\frac{\partial \overline{\rho} \widetilde{Z^{''2}}}{\partial t} + \frac{\partial \overline{\rho} \widetilde{u}_j \widetilde{Z^{''2}}}{\partial x_j} = \frac{\partial}{\partial x_j} \left[ \left( \overline{\rho} \widetilde{\alpha}_m+ \frac{\widetilde{\mu}_t}{Sc_t} \right) \frac{\partial \widetilde{Z^{''2}}}{\partial x_j} \right] - 2 \overline{\rho} \widetilde{\chi}_Z + 2 \frac{\widetilde{\mu}_t}{Sc_t} \frac{\partial \widetilde{Z}}{\partial x_j} \cdot \frac{\partial \widetilde{Z}}{\partial x_j } + \overline{\dot{S}}_{zv},
\end{aligned}
\end{equation}

\begin{equation}\label{c_eqn}
\begin{aligned}
&\frac{\partial \overline{\rho} \widetilde{c}}{\partial t} + \frac{\partial \overline{\rho} \widetilde{u}_j \widetilde{c}}{\partial x_j} = \frac{\partial}{\partial x_j} \left[ \left( \overline{\rho} \widetilde{\alpha}_m+ \frac{\widetilde{\mu}_t}{Sc_t} \right) \frac{\partial \widetilde{c}}{\partial x_j} \right] + \widetilde{\dot{\omega}_c} + \overline{\dot{S}}_{Y_{fuel}},
\end{aligned}
\end{equation}

\begin{equation}\label{cvar_eqn}
\begin{aligned}
&\frac{\partial \overline{\rho} \widetilde{c^{''2}}}{\partial t} + \frac{\partial \overline{\rho} \widetilde{u}_j \widetilde{c^{''2}}}{\partial x_j} = \frac{\partial}{\partial x_j} \left[ \left( \overline{\rho} \widetilde{\alpha}_m+ \frac{\widetilde{\mu}_t}{Sc_t} \right) \frac{\partial \widetilde{c^{''2}}}{\partial x_j} \right] - 2 \overline{\rho} \widetilde{\chi}_c + 2 \frac{\widetilde{\mu}_t}{Sc_t} \frac{\partial \widetilde{c}}{\partial x_j} \cdot \frac{\partial \widetilde{c}}{\partial x_j} + 2 \left( \widetilde{c \dot{\omega}_c} - \widetilde{c} \widetilde{\dot{\omega}_c} \right),
\end{aligned}
\end{equation}

\begin{equation}\label{zcvar_eqn}
\begin{aligned}
&\frac{\partial \overline{\rho} \widetilde{Z^{''} c^{''}}}{\partial t} + \frac{\partial \overline{\rho} \widetilde{u}_j \widetilde{Z^{''} c^{''}}}{\partial x_j} = \frac{\partial}{\partial x_j} \left[ \left( \overline{\rho} \widetilde{\alpha}_m+ \frac{\widetilde{\mu}_t}{Sc_t} \right) \frac{\partial \widetilde{Z^{''} c^{''}}}{\partial x_j} \right] - 2 \overline{\rho} \widetilde{\chi}_{Zc} + 2 \frac{\widetilde{\mu}_t}{Sc_t} \frac{\partial \widetilde{Z}}{\partial x_j} \cdot \frac{\partial \widetilde{c}}{\partial x_j} + \left( \widetilde{Z \dot{\omega}_c} - \widetilde{Z} \widetilde{\dot{\omega}_c} \right),
\end{aligned}
\end{equation}
where $\widetilde{\alpha}_m$ is the molecular diffusivity modeled as $\widetilde{\alpha}_m = \widetilde{\nu} / Sc$, $Sc = 1$ is the Schmidt number, $Sc_t = 0.4$ is the turbulent Schmidt number, $\mu_t$ is the turbulent dynamic viscosity, and $\dot{\omega}_c = \dot{\omega}_{H_2} + \dot{\omega}_{H_2 O} + \dot{\omega}_{CO} + \dot{\omega}_{CO_2}$ is the reaction rate for $c$. $\overline{\dot{S}}_{Y_{fuel}}$ is the production rate of fuel species because of evaporation. Since $\widetilde{c}$ is represented by reaction products, i.e., Eq.~(\ref{c_represent}), $\overline{\dot{S}}_{Y_{fuel}}$ is zero. The high-order source term $\overline{\dot{S}}_{zv}$ is hard to close without DNS data~\cite{pera2006modeling, Meftah2010SGSAO}, thus assumed to be zero. The terms $\widetilde{c \dot{\omega}_c}$ and $\widetilde{Z \dot{\omega}_c}$ are derived from the reaction source terms $\widetilde{c^{''} \dot{\omega}_c^{''}}$ and $\widetilde{Z^{''} \dot{\omega}_c^{''}}$ respectively~\cite{chen2015simulation}. 

For the convenience of joint PDF integration, the progress variable is normalized as $\widetilde{c}_n = \widetilde{Y}_c / \widetilde{Y}_{c,max}$ in the lookup table~\cite{chen2020prediction}, where $\widetilde{Y}_{c,max}$ is the maximum of $\widetilde{Y}_c$, obtained in a laminar premixed flame at a given $\widetilde{Z}$. The variances and covariances are normalized as $\widetilde{g}_Z = \widetilde{Z^{''2}} \Big/ \left[ \widetilde{Z} \left( 1 - \widetilde{Z} \right) \right]$, $\widetilde{g}_c = \widetilde{c^{''2}} \Big/ \left[ \widetilde{c} \left( \widetilde{Y}_{c,max} - \widetilde{c} \right) \right]$ and $\widetilde{g}_{cz}=\widetilde{Z^{''} c^{''}} \big/ \sqrt{\widetilde{Z^{''2}} \widetilde{c^{''2}}}$. Note that $\widetilde{g}_{cz}$ is the scalar correlation coefficient. Specific enthalpy reduction $\widetilde{h}_{r} = \widetilde{h}_{ad} - \widetilde{h}$, where $\widetilde{h}_{ad}$ is the adiabatic enthalpy.

Following Refs.~\cite{ruan2012scalar, kolla2009scalar, dunstan2013scalar, chen2017large}, the scaler dissipation rates $\widetilde{\chi}_Z$, $\widetilde{\chi}_c$ and $\widetilde{\chi}_{Zc}$ are modelled as
\begin{equation}\label{chi_Z_expr}
\begin{aligned}
&\overline{\rho} \widetilde{\chi}_Z = \overline{\rho \alpha_m \left( \frac{\partial Z^{''}}{\partial x_j} \frac{\partial Z^{''}}{\partial x_j} \right) } 
\simeq C_Z \overline{\rho} \left( \frac{\widetilde{\nu_t}}{\Delta^2} \right) \widetilde{Z^{''2}},
\end{aligned}
\end{equation}

\begin{equation}\label{chi_c_expr}
\begin{aligned}
&\overline{\rho} \widetilde{\chi}_c = \overline{\rho \alpha_m \left( \frac{\partial c^{''}}{\partial x_j} \frac{\partial c^{''}}{\partial x_j} \right) } 
\simeq \left[1 - \exp \left( -0.75 \frac{\Delta}{\delta_L^0} \right) \right] \left[ 2K_c^* \frac{S_L^0}{\delta_L^0} + \left( C_3 - \tau_{hr} C_4 Da_{\Delta}  \right) \frac{2 u'_\Delta}{3\Delta} \right] \frac{\widetilde{c^{''2}}}{\beta_c},
\end{aligned}
\end{equation}

\begin{equation}\label{chi_Zc_expr}
\begin{aligned}
&\overline{\rho} \widetilde{\chi}_{Zc} = \overline{\rho \alpha_m \left( \frac{\partial Z^{''}}{\partial x_j} \frac{\partial c^{''}}{\partial x_j} \right) } 
\simeq C_Z \overline{\rho} \left( \frac{\widetilde{\nu_t}}{\Delta^2} \right) \widetilde{Z^{''}c^{''}},
\end{aligned}
\end{equation}
where $C_Z=2$~\cite{pitsch2006large}, $\Delta$ denotes the filter width, $\nu_t$ is the turbulent kinetic viscosity, $u'_{\Delta} = |\hat{\tilde{\mathbf{u}}} - \tilde{\mathbf{u}}|$ is the SGS velocity, $C_3 = 1.5 \sqrt{Ka_{\Delta}} / ( 1 + \sqrt{Ka_{\Delta}} )$, $C_4 = 1.1 / (1 + Ka_{\Delta})^{0.4}$, the SGS Damk\"{o}hler number $Da_{\Delta} = (S_L^0 \Delta) / (u'_{\Delta} \delta_{L}^0)$, the SGS Karlovitz number $Ka_{\Delta} = (u'_{\Delta})^{1.5} (\Delta / \delta_L^0)^{-0.5}$, and the coefficient $\beta_c$ is taken to be 7.5 for simplicity~\cite{chen2017large}. The laminar flame speed $S_L^0$, laminar flame thickness $\delta_L^0$, heat release parameter $\tau_{hr} = (T_{ad} - T_{u}) / T_{u}$ and model parameter $K_c^*$~\cite{kolla2009scalar} are obtained from unstrained planar laminar premixed flame calculation~\cite{dunstan2013scalar}, where $T_{u}$ and $T_{ad}$ are the unburnt and adiabatic flame temperatures, respectively.

The aforementioned Favre-filtered variables, i.e., $\widetilde{\nu}$, $\widetilde{\kappa}$, $\widetilde{C}_p$, $\Delta \widetilde{h}_f^o$, $\widetilde{C}_{p,\text{eff}}$, $\widetilde{W}_{mix}$, $\widetilde{Y}_{c,max}$, $\widetilde{\dot{\omega}_c}$, $\widetilde{c \dot{\omega}_c}$ and $\widetilde{Z \dot{\omega}_c}$, are supposed to be initialized and updated in the time advancing of governing equations. Moreover, the mass fraction of ethanol, $\widetilde{Y}_{C_2 H_5 OH}$, is in demand for the calculation of evaporation rates, as shown in Section~\ref{liquid_phase_model}. It is a common practice to tabulate those physical quantities directly, since the mass fractions of all the species in large chemical mechanisms are too much to be stored in the lookup table. The laminar flamelets are obtained from the laminar premixed flame calculation, and then a proper representation of the joint PDF function is required to account for turbulent mixing and diffusion. In this work, the joint presumed PDF approach is applied.

\subsection{Joint PDF method with correlation}\label{jPDF_method}

In the FGM model, the flamelet/progress variable (FPV) model~\cite{pierce2004progress} and the conditional moment closure (CMC) model~\cite{ukai2013cmc}, the subgrid TCI is typically modeled using the presumed PDF method~\cite{jaganath2021transported}, as it offers lower computational cost compared to the transported PDF method and is straightforward to implement within existing computational fluid dynamics (CFD) codes~\cite{darbyshire2012presumed}. In the joint PDF method, the averaged value of a generic quantity $\Phi$ is modeled as
\begin{equation}\label{jPDF-general}
\begin{aligned}
&\widetilde{\Phi} = \int_{0}^{1} \int_{0}^{1} \Phi (\xi, \zeta) \mathcal{P}(\xi,\zeta) d\xi d\zeta,
\end{aligned}
\end{equation}
where $\xi$ and $\zeta$ are the sample space variable for $\widetilde{Z}$ and $\widetilde{c_n}$ respectively, $\Phi (\xi, \zeta)$ is obtained from the laminar flamelets, and $\mathcal{P}(\xi,\zeta)$ is the joint PDF of $\xi$ and $\zeta$. It is noticed that $\mathcal{P}(\xi,\zeta)$ is a bivariate distribution of statistically dependent variables. In probability theory and statistics, Sklar's theorem~\cite{Nelsen2006copulas} demonstrates that a multivariate joint distribution can be written in terms of ($\romannumeral1$) univariate marginal distribution functions that describe the randomness of variables, and ($\romannumeral2$) a \textit{copula} that describes the dependence structure among the random variables. A copula is a multivariate cumulative distribution function (CDF) $\mathcal{C}[F_1(X_1), F_2(X_2), \ldots, F_N(X_N)]$, for which the marginal CDFs $F_1(X_1), F_2(X_2), \ldots, F_N(X_N)$ of random variables $X_1, X_2, \ldots, X_N$ are uniformly distributed on the interval $[0,1]$~\cite{Nelsen2006copulas}. 
Under consideration in this study is the Frank copula~\cite{frank1979simultaneous}, which is a widely-used elliptically contoured Archimedean copula, written as
\begin{equation}\label{frank_copula}
\begin{aligned}
&\mathcal{C} = 
\begin{cases}
- \frac{1}{\theta} \log \left[ 1 + \frac{(e^{-\theta F_1} - 1) (e^{-\theta F_2} - 1)}{e^{-\theta} - 1} \right],& (\theta \in \mathbb{R} / \{ 0 \}),\\
F_1 F_2,& (\theta = 0).
\end{cases}
\end{aligned}
\end{equation}
If $\xi$ and $\zeta$ are independent, $\theta = 0$. Otherwise, $\theta$ is obtained via solving~\cite{Nelsen2006copulas}
\begin{equation}\label{frank_copula_theta}
\begin{aligned}
&1 + \frac{4}{\theta} \left( \int_{0}^{\theta} \frac{\varsigma}{e^{\varsigma} - 1} d\varsigma -1 \right) = \frac{6}{\pi} \arcsin \left( \frac{\widetilde{g}_{cz}}{2} \right).
\end{aligned}
\end{equation}

The commonly used marginal PDF in turbulent combustion for $F_1$ and $F_2$ is the $\beta$-distribution,
\begin{equation}\label{beta_pdf}
\begin{aligned}
&\mathcal{P}_{\beta} (\xi;a,b) = \frac{\Gamma (a + b)}{\Gamma (a) + \Gamma (b)} \xi^{a-1} (1-\xi)^{b-1},
\end{aligned}
\end{equation}
where $\Gamma$ is the Gamma function, and $a$ and $b$ are related to the mean $\widetilde{\xi}$ and variance $\widetilde{\xi^{''2}}$~\cite{ruan2014modelling}.

Now that $\mathcal{P}(\xi,\zeta)$ is described using a copula, Eq.~(\ref{jPDF-general}) can be derived as
\begin{equation}\label{jPDF-general-copula}
\begin{aligned}
&\widetilde{\Phi} = \int_{0}^{1} \int_{0}^{1} \Phi (\xi, \zeta) \frac{\partial^2 \mathcal{C} (F_1, F_2)}{\partial \xi \partial \zeta} d\xi d\zeta.
\end{aligned}
\end{equation}
The joint PDF in Ref.~\cite{darbyshire2012presumed, ruan2014modelling} is obtained by differentiating $\mathcal{C}$ in Eq.~(\ref{jPDF-general-copula}) with respect to $F_1$ and $F_2$, written as,
\begin{equation}\label{jPDF-copula-ruan}
\begin{aligned}
\widetilde{\Phi} = & \int_{0}^{1} \int_{0}^{1} \Phi \frac{\partial ^2 \mathcal{C}}{\partial F_1 \partial F_2} \mathcal{P}_{\beta}(\xi) \mathcal{P}_{\beta}(\zeta) d\xi d\zeta.
\end{aligned}
\end{equation}
Ref.~\cite{ruan2014modelling} has reported that the number of sampling points in $\xi$ and $\zeta$ directions should be as large as possible to ensure the statistical convergence of Eq.~(\ref{jPDF-copula-ruan}). The reason is that the gradients of $\mathcal{P}_{\beta}(\xi; a,b)$ are relatively large in the following cases: ($\romannumeral1$) $\mathcal{P}_{\beta}(\xi=0) \rightarrow \infty$ and $\mathcal{P}_{\beta}(\xi=1) \rightarrow \infty$ if $a<1$ and $b<1$, corresponding to $\widetilde{g}_Z > \widetilde{\xi} / (1 + \widetilde{\xi})$ and $\widetilde{g}_Z > (1 - \widetilde{\xi}) /  (2 - \widetilde{\xi})$; ($\romannumeral2$) $\mathcal{P}_{\beta}$ exhibits a sharp peak within a very narrow domain of $\xi$ if $a \gg b$ or $a \ll b$, i.e., if $\widetilde{g}_Z \ll 1$ and $\widetilde{\xi} \neq 1/2$. In the flow field, the second case is frequently encountered, since the fluctuations of mixture fraction are relatively low. Therefore, a considerable number of sampling points are required to ensure the accuracy of numerical integration in Eq.~(\ref{jPDF-copula-ruan}). 

Alternatively, Eq.~(\ref{jPDF-general-copula}) can be replaced by one that is easy to evaluate via integration by parts~\cite{langella2016large}, because the copula $\mathcal{C}$ changes slower than $\mathcal{P}_{\beta}$ in the $\xi$ and $\zeta$ directions. In this study, the joint PDF is derived by differentiating $\Phi$ with respect to $\xi$ and $\zeta$, as
\begin{equation}\label{jPDF-copula-new}
\begin{aligned}
\widetilde{\Phi} = & \int_{0}^{1} \int_{0}^{1} \frac{\partial ^2 \Phi}{\partial \xi \partial \zeta} \mathcal{C} d \xi d \zeta
-
\int_{0}^{1} \left( \frac{\partial \Phi}{\partial \zeta} \mathcal{C} \Bigg | _{\xi = 0}^{\xi = 1} \right) d\zeta
-
\int_{0}^{1} \left( \frac{\partial \Phi}{\partial \xi} \mathcal{C} \Bigg | _{\zeta = 0}^{\zeta = 1} \right) d\xi
+
\Phi \mathcal{C} \Bigg | _{\xi = 0}^{\xi = 1} \Bigg | _{\zeta = 0}^{\zeta = 1}.
\end{aligned}
\end{equation}

\begin{figure}[htb]
  \centering
    \includegraphics[scale=0.53]{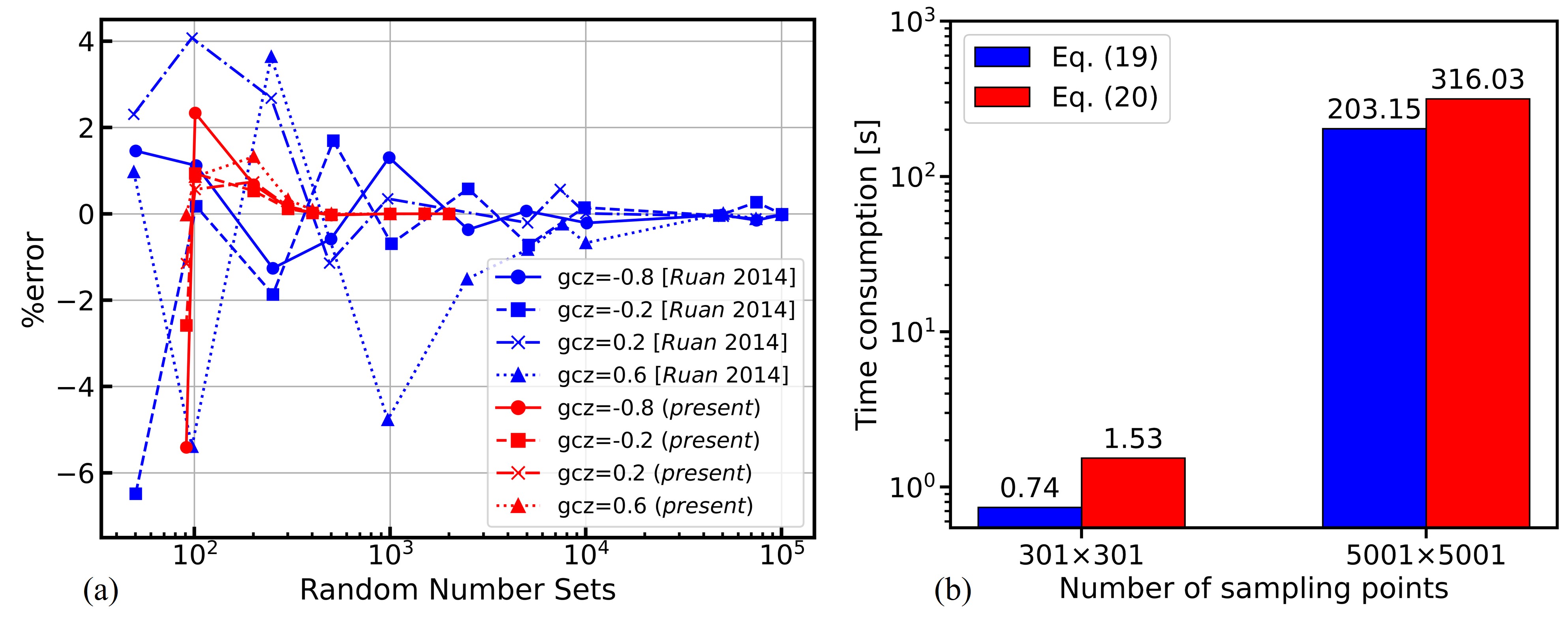}
    \caption{Numerical performance of Eq.~(\ref{jPDF-copula-ruan}) and Eq.~(\ref{jPDF-copula-new}) in computing the joint PDF, using $\widetilde{Z}=0.03$, $\widetilde{c}=0.7$, $\widetilde{Z^{''2}}=2.91\times 10^{-4}$ and $\widetilde{c^{''2}}=0.105$: (a) errors of reaction rate $\widetilde{\dot{\omega}_c}$ concerning the grid sizes; (b) time consumption for eleven $\widetilde{g}_{cz}$ points.}\label{fig1_sample}
\end{figure}

The advantages of Eq.~(\ref{jPDF-copula-new}) comparing with Eq.~(\ref{jPDF-copula-ruan}) are: ($\romannumeral1$) rapid grid convergence, which notably saves computational time; and ($\romannumeral2$) good convenience to use copulas with implicit formulas, such as the Gaussian copula. Ref.~\cite{ruan2014modelling} has reported the relative errors in $\widetilde{\dot{\omega}_c}$ against the sample grid sizes in ($\xi$, $\zeta$) space, which provides a benchmark for numerical accuracy analysis. The test case is the stoichiometric unstretched hydrogen/air combustion, using a chemical kinetics mechanism of 11 species and 25 reactions~\cite{li2004updated}. Convergence criteria is that the error is less than 1\% comparing the saturation mesh resolution. Fig.~\ref{fig1_sample} presents the numerical errors and time consumption using Eq.~(\ref{jPDF-copula-ruan}) and Eq.~(\ref{jPDF-copula-new}). The numerical differentiation method is the five-point formula, while the integration method is the trapezoidal rule. The errors of Eq.~(\ref{jPDF-copula-new}) is less than 1\% in the case using $301 \times 301$ samples points for various correlation coefficients $\widetilde{g}_{cz}$, compared with $5000 \times 5000$ samples~\cite{ruan2014modelling} required for Eq.~(\ref{jPDF-copula-ruan}). Similar behaviors are observed for other equivalence ratios. Fig.~\ref{fig1_sample}(b) shows the computational time measured on the $\widetilde{g}_{cz}$ grid with 11 equally spaced points over $[-1,1]$. Although Eq.~(\ref{jPDF-copula-new}) introduces some computational complexity when applied to the same random sample sets as Eq.~(\ref{jPDF-copula-ruan}), it achieves a speedup of over two orders of magnitude owing to its rapid convergence.

\begin{figure}[htb]
  \centering
    \includegraphics[scale=0.4]{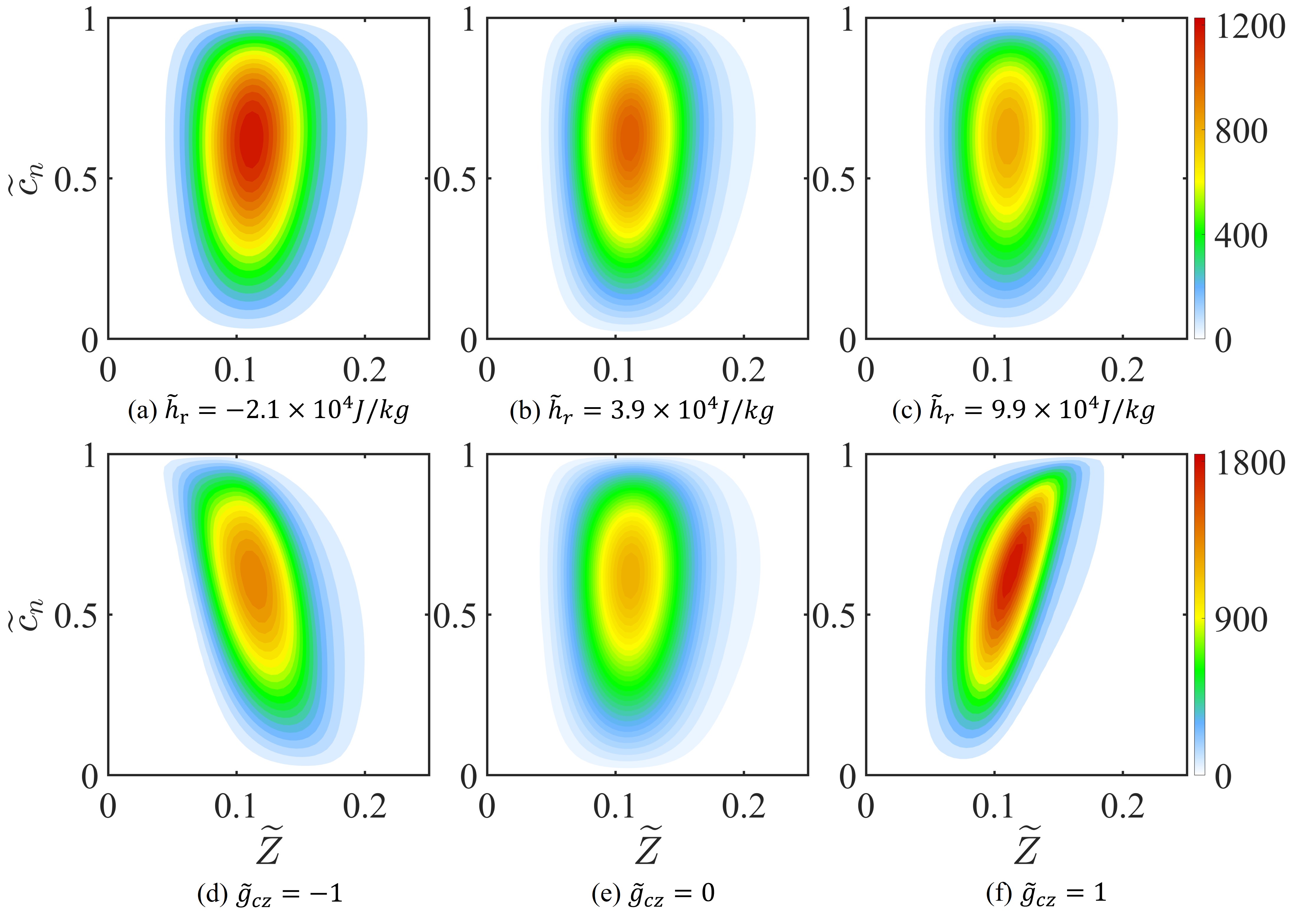}
    \caption{Contour plot of the reaction rate $\widetilde{\dot{\omega}_c}$ with respect to: (a)-(c) the evaporation-induced specific enthalpy reduction $\widetilde{h}_{r}$; and (d)-(f) the correlation coefficient $\widetilde{g}_{cz}$. Parameters $\widetilde{g}_Z=4.1\times 10^{-3}$ and $\widetilde{g}_c=0.4$ are chosen for visualization. }\label{fig2_omegac_tab}
\end{figure}

Fig.~\ref{fig2_omegac_tab} shows the effects of specific enthalpy reduction $\widetilde{h}_{r}$ and scalar correlation efficient $\widetilde{g}_{cz}$ on the reaction rate $\widetilde{\dot{\omega}_c}$ in the $\widetilde{Z} \times \widetilde{c}_n$ space. Intermediate values of $\widetilde{g}_Z=4.1\times 10^{-3}$ and $\widetilde{g}_c=0.4$ are used for illustration. Figs.~\ref{fig2_omegac_tab}(a) and (e) are the same data using different color bars. Figs.~\ref{fig2_omegac_tab}(a)-(c) illustrate that the increase of $\widetilde{h}_{r}$ could decelerate the chemical reactions and narrow down the range of flammability limits. Figs.~\ref{fig2_omegac_tab}(d)-(f) demonstrate that a positive or negative value of the scalar correlation coefficient $\widetilde{g}_{cz}$ not only results in a positive or negative slope on the reaction rate contour in the $\widetilde{Z} \times \widetilde{c}_n$ directions, but also increases the maximum value of $\widetilde{\dot{\omega}_c}$. Such influences of $\widetilde{g}_{cz}$ on $\widetilde{\dot{\omega}_c}$ are generally in good agreement with the DNS analysis reported in Ref.~\cite{chen2018priori}.

\subsection{Liquid phase modeling}\label{liquid_phase_model}

The Lagrangian method is used to track a large number of liquid parcels, each of which represents a finite number of dispersed spherical droplets. To be solved in the Lagrangian framework are the evolution equations for droplet properties, including the droplet position $\mathbf{x}_d$, mass $m_d$, velocity $\mathbf{u}_d$ and temperature $T_d$, where the subscript $d$ refers to the property of a droplet. The stochastic collision among the droplets can be neglected, since the targeted cases EtF1, EtF4 and EtF7~\cite{GOUNDER20123372}, are dilute spray flames, in which the volume fractions of the dispersed phase are $3.3\times 10^{-4}$, $1.1\times 10^{-4}$ and $2.5\times 10^{-4}$ respectively, lower than $1\times 10^{-3}$~\cite{michaelides2022multiphase}. Given that the density of liquid-phase ethanol is much larger than that of the gas mixture, the considered forces acting on the droplets are the gravity and drag force~\cite{huspray2017}. The governing equations of displacement, mass, momentum and energy for each droplet are written as~\cite{de2013large, huspray2017}
\begin{equation}\label{droplet_displacement}
\begin{aligned}
\frac{d \mathbf{x}_d}{dt} = \mathbf{u}_d,
\end{aligned}
\end{equation}

\begin{equation}\label{droplet_mass}
\begin{aligned}
\frac{d m_d}{d t} = \dot{m}_d = -2 \pi r_d \rho \hat{D} Sh \ln (1+B_M),
\end{aligned}
\end{equation}

\begin{equation}\label{droplet_momentum}
\begin{aligned}
\frac{d \mathbf{u}_d}{dt} = \frac{3}{8r_d} \frac{\rho}{\rho_d} C_D \left( \mathbf{u} - \mathbf{u}_d \right) |\mathbf{u} - \mathbf{u}_d| + \left( 1 - \frac{\rho}{\rho_d} \right)\mathbf{g},
\end{aligned}
\end{equation}

\begin{equation}\label{droplet_energy}
\begin{aligned}
\frac{d T_d}{d t} = \frac{2 \pi r_d \kappa Nu }{m_d C_{p,l}} \frac{B_H (T - T_d)}{e^{B_H} - 1} + \frac{\dot{m}_d}{m_d} \frac{h_v}{C_{p,l}},
\end{aligned}
\end{equation}
where $m_d=4/3 \pi \rho_l r_d^3$ is the mass of the droplet, $\rho_l$ is the liquid density, $\hat{D}$ is the vapour mass diffusivity in the gas mixture, $\mathbf{u}$ denotes the gas-phase velocity, $\mathbf{g}$ corresponds to the gravitational acceleration, $C_{p,l}$ is the heat capacity of liquid phase, and $h_v$ is the latent heat of evaporation.

In Eq.~(\ref{droplet_mass}), the Spalding numbers of heat and mass transfer are written as~\cite{de2013large, palanti2019implicit}
\begin{equation}\label{droplet_spalding}
\begin{aligned}
B_H = C_p \frac{T - T_d }{h_v},
~B_M= \frac{X_{Fs}-X_{\infty}}{1-X_{Fs}},
\end{aligned}
\end{equation}
where $X_{\infty}$ and $X_{Fs}$ are the mole fractions of vapor fuel in the local ambient gas mixture and at the droplet surface, respectively. Considering the thermodynamic non-equilibrium effect using the Langmuir-Knudsen law~\cite{miller1998evaluation}, $X_{Fs}$ is calculated as
\begin{equation}\label{droplet_X_Fs}
\begin{aligned}
&X_{Fs} = X_{F} \frac{p_{Sat}}{p}\exp{\left[ \frac{h_v W_v}{R_0} \left( \frac{1}{T_b} - \frac{1}{T_d} \right) \right]} - \beta_v \frac{L_K}{r_d},
\end{aligned}
\end{equation}
where $X_{F}$ is the mole fraction of vapor fuel in the droplet, $p_{Sat}$ is the saturation pressure, $W_v$ is the molecular weight of vapor fuel, and $T_b$ is the temperature of the liquid fuel at the boiling point. The Knudsen layer thickness $L_K$ and non-dimensional parameter $\beta_v$ are calculated as~\cite{miller1998evaluation}
\begin{equation}\label{droplet_L_K_beta_v}
\begin{aligned}
L_K = \frac{\mu \sqrt{2\pi T_d R_0 / W_v}}{p Sc},~\beta_v = \frac{-\dot{m}_d Pr}{4\pi r_d \mu},
\end{aligned}
\end{equation}
where $Pr$ is the gas-phase Prandtl number. The Sherwood number $Sh = 2 + 0.6 Re_d^{1/2} Sc^{1/3}$~\cite{ranz1952evaporation}, where $Sc$ is the Schmidt number in the gas phase, and the droplet Reynolds number $Re_d \equiv 2 \rho r_d |\mathbf{u}_d - \mathbf{u}| / \mu$ is computed via the difference of velocity between two phases, with $\mu$ as the gas-phase dynamic viscosity.

In Eq.~(\ref{droplet_momentum}), the particle drag coefficient $C_d$ is modeled as~\cite{putnam1961integratable}
\begin{equation}\label{droplet_Cd}
\begin{aligned}
C_d = 
\begin{cases}
\frac{24}{Re_d} \left( 1 + \frac{1}{6} Re_d^{2/3} \right),& (Re_d \leq 1000),\\
0.424,& (Re_d > 1000).
\end{cases}
\end{aligned}
\end{equation}

In Eq.~(\ref{droplet_energy}), the Nusselt number $Nu= 2 + 0.6 Re_d^{1/2} Pr^{1/3}$ is approximated using the correlation by Ranz and Marshall~\cite{ranz1952evaporation}.

The spray-related source terms in Eqs.~(\ref{mass_eqn})-(\ref{enthalpy_eqn}), i.e., $\overline{\dot{S}_v}$, $\overline{\dot{\mathbf{S}}}_{m}$ and $\overline{\dot{S}}_{h}$, are obtained by
\begin{equation}\label{droplet_source_Sv}
\begin{aligned}
& \overline{\dot{S}_v} = - \frac{1}{V_c} \sum_{k=1}^{n_d} N_{d,k} \dot{m}_{d,k},
\end{aligned}
\end{equation}

\begin{equation}\label{droplet_source_S_m}
\begin{aligned}
& \overline{\dot{\mathbf{S}}_{m}} = - \frac{1}{V_c} \sum_{k=1}^{n_d} N_{d,k} \left( m_{d,k} \frac{d \mathbf{u}_{d,k}}{dt} + \dot{m}_{d,k} \mathbf{u}_{d,k} \right),
\end{aligned}
\end{equation}

\begin{equation}\label{droplet_source_Sh}
\begin{aligned}
& \overline{\dot{S}_h} = \frac{1}{V_c} \sum_{k=1}^{n_d} N_{d,k} \left[ 2\pi r_d \kappa Nu (T_d - T) - \dot{m}_{d,k} h(T_{d,k})  \right],
\end{aligned}
\end{equation}
where $n_d$ is the total number of parcels passing through the grid cell under consideration, $N_{d,k}$ is the number of particles in the parcel $k$, $V_c$ is the cell volume, and $h(T_d)$ is the vapor enthalpy at the droplet temperature $T_d$.

\section{Experiments and numerical details}\label{sec_experiments}

\begin{figure}[htb]
  \centering
    \includegraphics[scale=0.5]{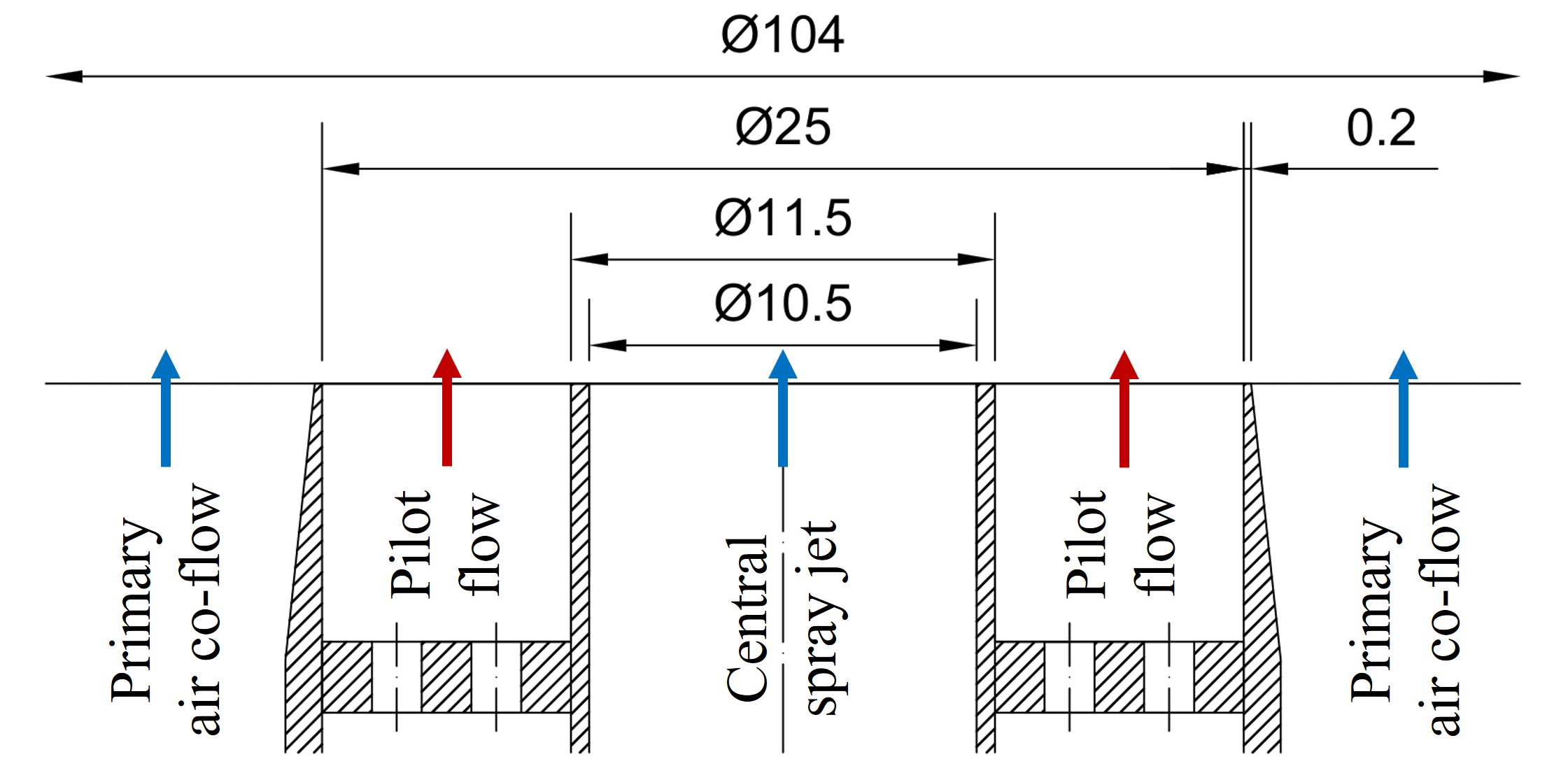}
    \caption{Schematic of the jet nozzle exit in the experiments~\cite{Gounder2009PhD, GOUNDER20123372}.}\label{fig0_sydneySpray_schemetric}
\end{figure}

The flame configuration experimentally studied at the University of Sydney~\cite{Gounder2009PhD, GOUNDER20123372} is investigated in this work. There are three incoming streams to the spray combustor, i.e., the central fuel jet, the pilot stream, and the air co-flow, as sketched in Fig.~\ref{fig0_sydneySpray_schemetric}. The central jet carries liquid and vapor fuel in a diameter of $d_{jet}=10.5$mm. The annular hot pilot flow at a velocity of 11.6m/s is injected to stabilize the flame, comprising a burnt stoichiometric mixture of acetylene, hydrogen and air. The air co-flow is provided using a wind tunnel at a velocity of 4.5m/s. Ethanol spray is well atomized using an ultrasonic nebulizer, avoiding numerical modeling for atomization and secondary breakup in the chamber. Radial profiles of mean droplet axial velocity, droplet axial root mean square (RMS) velocity fluctuations, normalized droplet volume fraction and droplet size distribution, are measured at $x/d_{jet}=0.3, 10, 20, 30$ from the nozzle exit in the Experiment A data set~\cite{GOUNDER20123372}. Radial profiles of the mean gas-phase temperature are measured at the axial locations of $x/d_{jet}=10, 20, 30$. The mean droplet radial velocity and radial RMS velocity fluctuations at the nozzle exit are reported in the Experiment B data set~\cite{GOUNDER20123372} which has slightly more percentages of vapor fuel flow rate than Experiment A. The present work studies three ethanol spray flames, i.e., EtF1, EtF4 and EtF7~\cite{GOUNDER20123372}, which mark the extremes in the experiments, as shown in Table~\ref{tbl_etfcases}. The EtF1 and EtF4 cases are designed to study the effects of increasing the liquid fuel flow rate for a fixed carrier velocity, while the EtF1 and EtF7 flames are used to investigate the effects of increasing the carrier velocity at a fixed liquid flow rate~\cite{GOUNDER20123372}.

\begin{table}[htb]
\caption{Experimental conditions of the ethanol spray jets, EtF1, EtF4 and EtF7~\cite{GOUNDER20123372}.}\label{tbl_etfcases}
\centerline{\begin{tabular}{l c c c}
\hline
Case ID                                    &EtF1     &EtF4     &Etf7   \\
\hline
Bulk jet velocity (m/s)                    &24       &24       &60     \\
Carrier (air) mass flow rate (g/min)       &150      &150      &376    \\
Liquid fuel injection rate (g/min)         &75       &23.4     &75     \\
Liquid flow rate at jet exit (g/min)       &45.7     &14.5     &73.0   \\ 
Vapor fuel flow rate at jet exit (g/min)   &29.3     &8.9      &2.0    \\ 
Mixture fraction at jet exit               &0.1636   &0.056    &0.0053 \\
Gas-phase temperature at jet exit (K)      &304.5    &280      &298    \\
Jet Reynolds number                        &22,200   &17,500   &45,700 \\
\hline
\end{tabular}}
\end{table}

The LES simulations are performed using the dfSprayFoam solver of the open-source platform DeepFlame~\cite{mao2024integrated} based on OpenFOAM libraries. The computational domain extends to $40d_{jet} \times 10d_{jet} \times 2\pi$ in cylindrical coordinates ($x,r,\Theta$), which is split to $501\times182\times64$ grid points in the axial, radial and azimuthal directions, respectively. A fine mesh is applied within the shear layer, near the nozzle exit and in the reaction zone. The viscosity ratio is smaller than approximately 20 ($r_v=\nu_t / \nu \leq 20$), corresponding to a sufficient resolution of resolving more than 80\% of kinetic energy~\cite{rittler2015sydney, pettit2011large}. The turbulent viscosity is approximated using the Sigma model, with a model parameter of $C_m=1.5$~\cite{rittler2015sydney}. The implicit Euler time-marching scheme is employed. Spatial gradients are evaluated via a second-order gradient scheme, while a second-order central difference scheme is adopted in the divergence terms. Simulations are carried out on 5 HPC nodes (140 CPU cores in total). Each node is equipped with two Intel$^\circledR$ Xeon$^\circledR$ Gold 6132 processors and 96GB memory. After the flows fully develop, statistics are collected over 10 flow-through times, which is sufficiently long for the convergence of liquid-phase statistics~\cite{de2013large}. 

The inlet conditions for the two phases are taken from experimental data at $x/d_{jet}=0.3$. Radial profiles of gas-phase mean axial and radial velocities are represented by velocities of droplets smaller than 10$\mu$m~\cite{GOUNDER20123372}. A synthetic eddy turbulent generator~\cite{kornev2007method} is imposed on the inlet velocities of the central fuel jet and air co-flow. The Reynolds normal stresses imposed on the central jet are given by the RMS velocity fluctuations measured in the experiments, while the Reynolds shear stresses are assigned to be zero. A uniform velocity profile of the co-flow at 4.5m/s is obtained at the nozzle exit plane with a relative turbulence intensity of about 5\%~\cite{Gounder2009PhD, GOUNDER20123372}. The radial profile of gas-phase mean axial velocity in the pilot flow at the jet exit plane is approximated using RANS simulations in a computational domain which further extends $50$mm upstream of the exit plane at the flow direction. The temperature of air co-flow is set as the ambient temperature of 298K~\cite{de2013large}, and the temperature of pilot flow is the adiabatic temperature of 2169K. Following Ref.~\cite{de2013large}, the drop axial velocity $u_{1,d}$ is provided using a Gaussian distribution, 
\begin{equation}\label{droplet_velocity_setup}
\begin{aligned}
&u_{1,d}(r)=\overline{u}_{1,d}(r) + u^{'}_{1,d}(r)\sqrt{2}\text{erf}^{-1}(2q_d-1),
\end{aligned}
\end{equation}
where $q_d$ is a random number from a uniform distribution in $(0,1)$, and $\overline{u}_{1,d}$ and $u^{'}_{1,d}$ are the droplet axial mean and RMS velocities measured in the experiments, respectively. The radial and azimuthal components of droplet velocity are given by the experimental results. The injected parcel diameters are assigned in the Rosin-Rammler distribution, consistent with the measurements~\cite{GOUNDER20123372}, as presented in Table~\ref{tbl_dropletsize}.

\begin{table}[htb]
\caption{Droplet Rosin-Rammler distributions in the central jet, measured at $x/d_{jet}=0.3$~\cite{GOUNDER20123372}.}\label{tbl_dropletsize}
\centerline{\begin{tabular}{l c c c}
\hline
Case ID                                 &EtF1      &EtF4     &Etf7   \\
\hline
Mean diameter ($\mu$m)                  &29        &32       &22     \\
Minimum diameter ($\mu$m)               &5         &3.5      &5      \\
Maximum diameter ($\mu$m)               &76        &77       &78     \\
Shape parameter                         &1.45      &1.6      &1.35   \\
\hline
\end{tabular}}
\end{table}

The lookup table for ethanol spray flames is computed using a detailed chemical kinetics mechanism with 54 species and 268 reactions~\cite{williams2014chemical}. The six-dimensional table has $25\times 81 \times 51 \times 15 \times 21 \times 11$ grids in the $\widetilde{h}_{r} \times \widetilde{Z} \times \widetilde{c}_n \times \widetilde{g}_Z \times \widetilde{g}_c \times \widetilde{g}_{cz}$ directions. An equidistant grid is employed in the $\widetilde{h}_{r}$ direction in the range of $[-2.1\times 10^4 , 9.9\times 10^4]$J/kg, where the maximum value corresponds to flame quenching. The grid in the $\widetilde{Z}$ direction is non-uniform in $[0,1]$, with refinement in the flammability limits. The grid along the $\widetilde{g}_Z$ direction is logarithmically distributed in the range of $[1\times 10^{-4},0.1]$, along with an additional point of $\widetilde{g}_Z=0$. The equidistant grids are utilized in the $\widetilde{c}_n$ and $\widetilde{g}_c$ directions over the interval $[0,1]$, as well as in the $\widetilde{g}_{Zc}$ direction within the range of [-1,1]. For the joint PDF method, $2725 \times 301$ equidistant grids in the $\xi \times \zeta$ directions are sufficient in the presence of two-phase heat transfer and evaporation. To load the high-dimensional table into memory, this study uses the Message Passing Interface (MPI) shared memory technique~\cite{hoefler2013mpi}. Further details are provided in the Supplementary Material.

For each flame, simulations with the following methods are carried out:
\begin{itemize}
\item FGM model considering scalar covariance and evaporation-induced specific enthalpy reduction due to evaporation (Tab6D).
\item FGM model considering evaporation-induced specific enthalpy reduction while neglecting effects of covariance, i.e., $\widetilde{g}_{cz}=0$ (Tab5D).
\item FGM model neglecting effects of scalar covariance and evaporation-induced specific enthalpy reduction, i.e., $\widetilde{g}_{cz}=0$ and $\widetilde{h}_{r}=0$ (Adiabatic).
\end{itemize}
Thus, Tab6D, Tab5D and Adiabatic utilize the lookup tables in 6, 5 and 4 dimensions, respectively, while sharing the same scalar transport equations, Eqs.~(\ref{z_eqn})-(\ref{zcvar_eqn}). The effects of scalar correlation on the spray flames are examined by comparing Tab6D and Tab5D, while the impact of specific enthalpy reduction is investigated through a comparison of Tab5D and Adiabatic.

\section{Results and discussion}\label{sec_results}

\subsection{Instantaneous fields}\label{sec_instantaneous}

\begin{figure}[htb]
  \centering
    \includegraphics[scale=0.45]{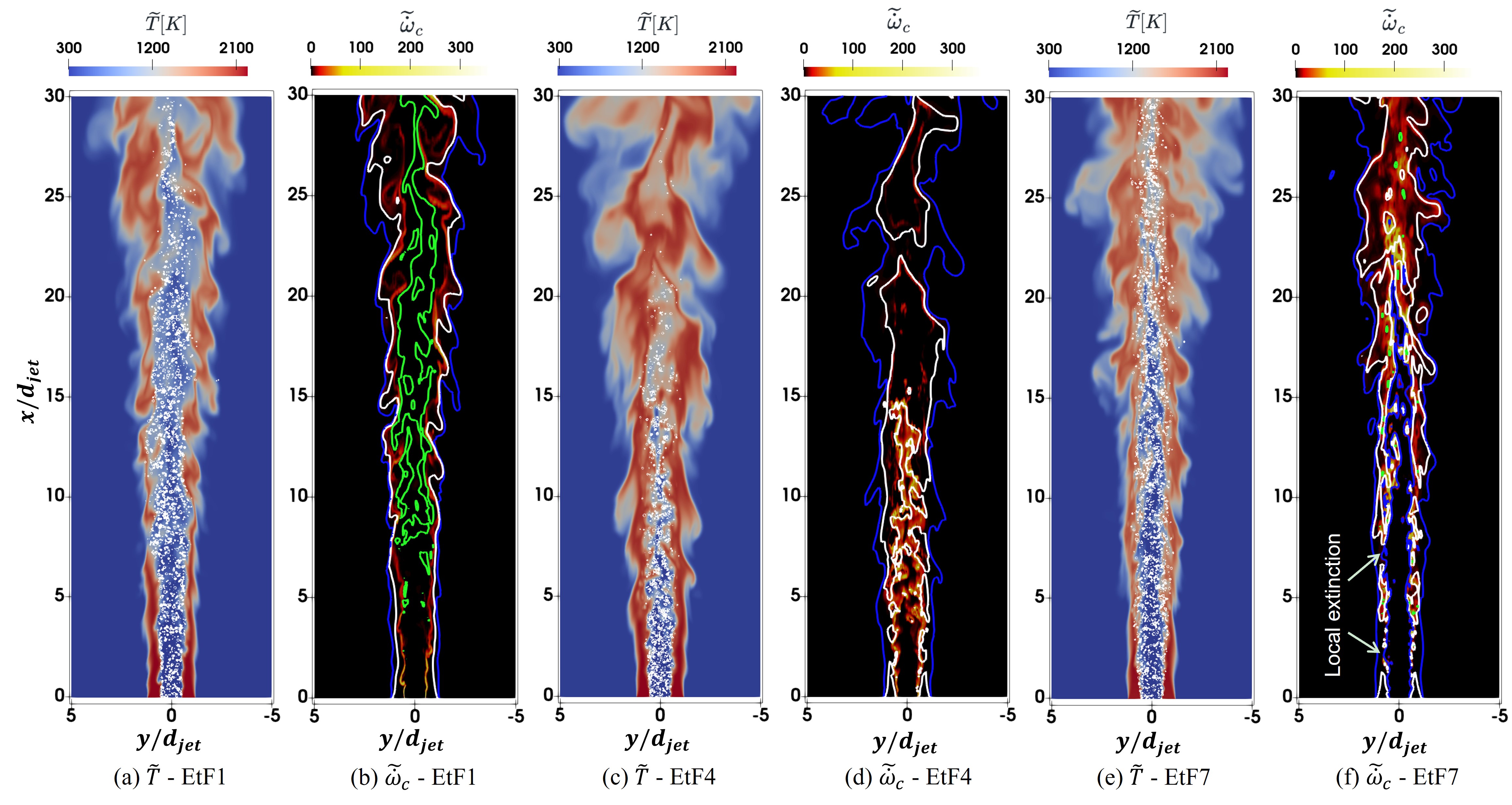}
    \caption{Instantaneous snapshots of temperature $\widetilde{T}$ and reaction rate $\widetilde{\dot{\omega}}_c$ in the EtF1, EtF4 and EtF7 flames. Contours in $\widetilde{\dot{\omega}}_c$ fields are plotted as: $\widetilde{Z}=0.047$ (blue), $\widetilde{Z}=\widetilde{Z}_{st}=0.1$ (white), and $\widetilde{Z}=0.27$ (green). Parcels are shown in $\widetilde{T}$ fields.}\label{fig3_instant_T_omega}
\end{figure}

\begin{figure}[htb]
  \centering
    \includegraphics[scale=0.43]{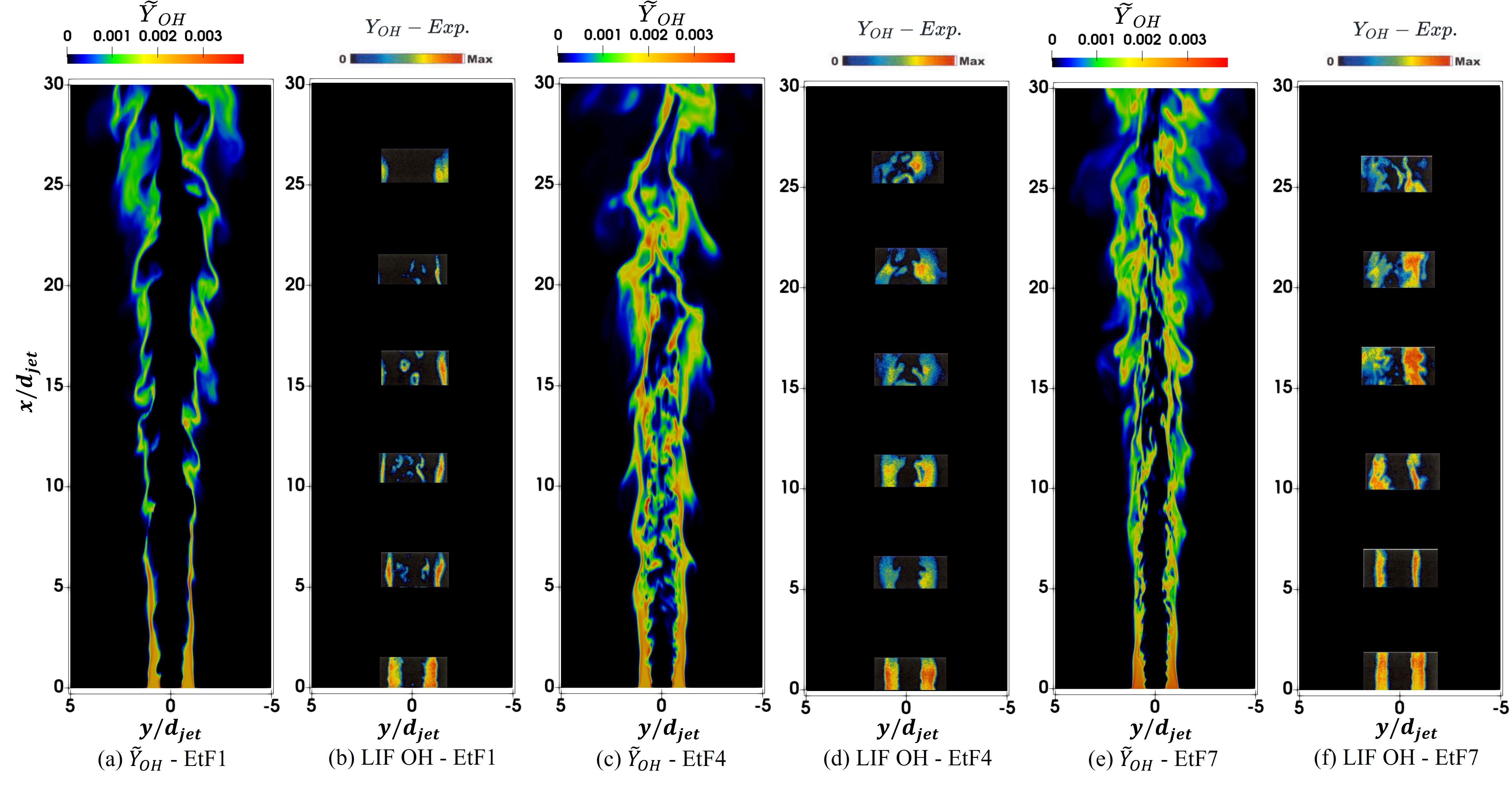}
    \caption{Instantaneous snapshots of OH mass fractions $\widetilde{Y}_{OH}$ in the EtF1 (a), EtF4 (c) and EtF7 (e) flames, compared to the LIF OH images (b,d,f) in the experiments~\cite{Gounder2009PhD}.}\label{fig4_instant_OH}
\end{figure}

Figure~\ref{fig3_instant_T_omega} presents instantaneous contours of the gas-phase temperature $\widetilde{T}$ and the chemical reaction rate $\widetilde{\dot{\omega}}_c$ for the EtF1, EtF4, and EtF7 flames, obtained using Tab6D. Contours in $\widetilde{\dot{\omega}}_c$ fields are plotted as $\widetilde{Z}=0.047, 0.1, 0.27$, corresponding to the lower flammability limit (LFL), stoichiometric mixture fraction and upper flammability limit (UFL), respectively. A comparison of EtF1 and EtF4 indicates that a higher fuel mass flow rate results in a longer flame. In Fig.~\ref{fig3_instant_T_omega}(b), the mean reaction zone ($\widetilde{\dot{\omega}}_c > 1$) in EtF1 does not penetrate the central jet region within the length of $x/d_{jet}=25$, which is consistent with Ref.~\cite{kirchmann2021two}. This is attributed to the high fuel concentration near or beyond the UFL. The EtF4 flame, characterized by the lowest mass flow rates of ethanol and its air carrier, has the shortest central reaction zone, approximately $x/d_{jet}=15$ in Fig.~\ref{fig3_instant_T_omega}(d), which is in accord with Ref.~\cite{kirchmann2021two}. The chemical reactions in EtF4 predominantly occur around the stoichiometric mixture fraction, leading to relatively high reaction rates. Because of the carrier velocity of 60m/s and the vapor fuel flow rate of 2g/min, EtF7 exhibits a high turbulence intensity and a long flame length in Fig.~\ref{fig3_instant_T_omega}(e), in accord with Refs.~\cite{hu2020large, huspray2017, kirchmann2021two}. In Fig.~\ref{fig3_instant_T_omega}(f), partial extinction is observed upstream of $x/d_{jet}=10$, due to the intense turbulence from the central jet. Along the axial direction, droplet evaporation gradually transforms the combustion from fuel-lean (about $x/d_{jet}=5$) to fuel-rich (about $x/d_{jet}=25$).

Fig.~\ref{fig4_instant_OH} presents instantaneous snapshots of the OH mass fraction $\widetilde{Y}_{OH}$, compared with the planar laser-induced fluorescence (LIF) OH images obtained in experiments~\cite{GOUNDER20123372}. The OH concentration provides insight into the location and shape of the burnt side of the flame front, as well as the flame length and the flame lift-off heights~\cite{rittler2015sydney}. Fig.~\ref{fig4_instant_OH} shows that the flame widths are well captured in the current simulations. At $x/d_{jet}=5$, the OH regions in EtF1 and EtF4 exhibit a broad, wrinkled structure, while in EtF7, strong turbulence and large vortex structures distort the OH region at $x/d_{jet}=15$. However, the inner secondary OH region in Fig.~\ref{fig4_instant_OH}(b), appearing as isolated, disconnected OH pockets, is not captured by the simulation in Fig.~\ref{fig4_instant_OH}(a). Gounder~\cite{Gounder2009PhD} attributed this secondary reaction zone to the pre-mixing of carrier air and the ethanol vapor, which forms a combustible mixture that ignites under favorable conditions. In the experiments, the inner OH region is not present in every LIF image~\cite{Gounder2009PhD}: ($\romannumeral1$) the frequency of occurrence of the secondary reaction zone increases with increasing fuel loading (in the direction of EtF4 $\rightarrow$ EtF3 $\rightarrow$ EtF1); ($\romannumeral2$) the secondary OH region is apparent with decreasing carrier velocity (in the direction of EtF7 $\rightarrow$ EtF5 $\rightarrow$ EtF2 $\rightarrow$ EtF1). 

One potential reason for the absence of the secondary OH region in EtF1 is the nonuniform mixing of gas-phase ethanol in the central jet. Some droplets, generated by the nebulizer, adhere to the pipe wall during transit to the jet exit, forming the liquid coating. When the air carrier velocity is fixed, the amount of liquid fuel adhering to the wall increases with fuel loading (from EtF4 to EtF1). Similarly, for a fixed ethanol mass fraction, the amount of adhered droplets increases as the mass flow rate of the air carrier decreases (from EtF7 to EtF1). These adhered droplets are subsequently blown to the jet exit plane, where they break up into liquid films, evaporate, and ignite due to the hot pilot flow. Therefore, the mixture fraction at $r/d_{jet}=0.5$ may be higher than that at $r/d_{jet}=0$. Downstream of $x/d_{jet}=5$, the shear forces acting on the gas mixture and the droplets cause them to separate radially, resulting in two distinct OH zones. In the experimental data, the fuel vapor mass flow rate at the jet exit plane is obtained as the difference between the injected and measured liquid flow rates~\cite{Gounder2009PhD, GOUNDER20123372}. However, the measurements did not account for the non-uniform distribution of $\widetilde{Z}$ in the central jet. Two potential indications can support this hypothesis: ($\romannumeral1$) the vapor fuel mass flow rate in EtF1 is sufficiently high that the saturated vapor pressure requires a mixture temperature exceeding $304K$, yet no additional pre-heating equipment was reported in Refs.~\cite{Gounder2009PhD, GOUNDER20123372}; and ($\romannumeral2$) the measured gas temperature in EtF1 at $r/d_{jet}=0$, $x/d_{jet}=10$ is similar to that of EtF4, whose combustion occurs mostly at the stoichiometric mixture fraction. Other evidence for this hypothesis will be discussed in Sections~\ref{sec_gasPhase} and \ref{sec_liquidPhase}. Additionally, it was noticed in Ref.~\cite{hu2020large} that the droplet size distribution measured at the jet exit plane exhibits distinct variations in different radial positions, with a pronounced bias toward smaller droplets as the measurement location approaches the pipe wall. But the attempt to apply the non-uniform fuel distribution in EtF7 was found to have little impact on the two-phase statistics~\cite{hu2020large}. The reason may be that the pre-vaporized gaseous fuel in EtF7 exhibits an extremely low concentration (0.5\%) in the total gaseous mixture of the central jet, while it is much higher in EtF1 (16.3\%). Moreover, the secondary OH region was not reported in acetone flames~\cite{Gounder2009PhD}, likely because the attached fuel already evaporates within the pipe due to acetone’s lower boiling point (329K compared to 351K for ethanol)~\cite{GOUNDER20123372}. 

\subsection{Gas phase statistics}\label{sec_gasPhase}

\begin{figure}[htb]
  \centering
    \includegraphics[scale=0.42]{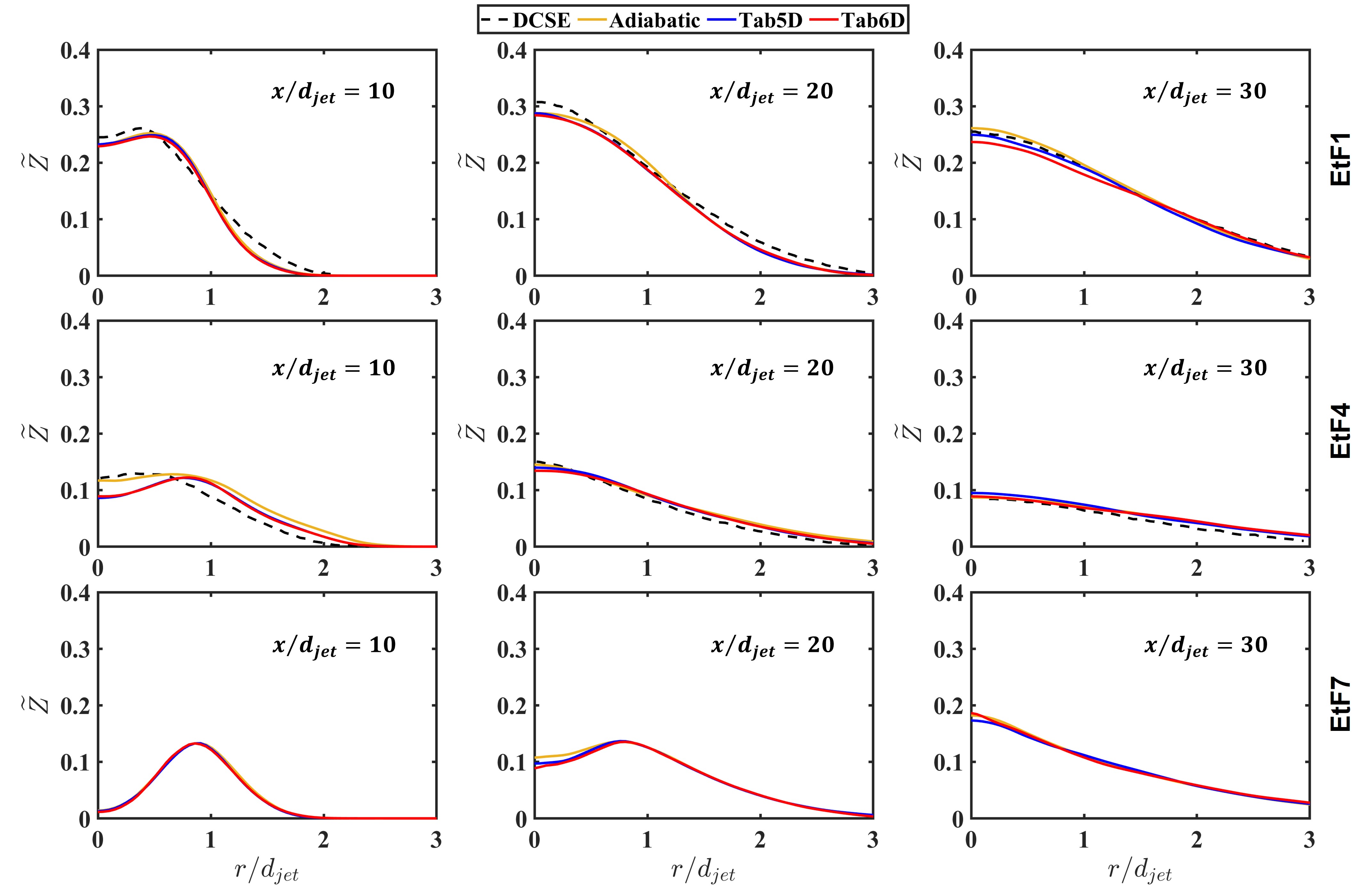}
    \caption{Radial profile of mean mixture fraction $\widetilde{Z}$ at different axial locations, $x/d_{jet}=10,20,30$, in the EtF1, EtF4 and EtF7 flames, compared with the DCSE results in Ref.~\cite{hussien2022simulations}. }\label{fig5_Z}
\end{figure}

No experimental data is available on the distribution of the Favre-averaged mixture fraction $\widetilde{Z}$. However, Ref.~\cite{hussien2022simulations} has reported the radial profile of $\widetilde{Z}$ for the EtF1, EtF3 and EtF4 flames, allowing for qualitative analysis. Fig.~\ref{fig5_Z} presents the radial profile of $\widetilde{Z}$ at different axial locations for the EtF1, EtF4 and EtF7 flames, compared to the Doubly Conditional Source-term Estimation (DCSE) results in Ref.~\cite{hussien2022simulations}, demonstrating good agreement. For EtF1, both FGM and DCSE predict a high concentration of ethanol along the axis of the combustion domain, which inhibits the formation of the secondary OH zone and may lead to an underprediction of gas-phase temperature. Therefore, the omission of $\widetilde{h}_{r}$ and $\widetilde{g}_{cz}$ in the lookup table has minimal impact on the $\widetilde{Z}$ distribution in EtF1. In EtF4, Adiabatic predicts a higher $\widetilde{Z}$ than Tab5D and Tab6D at $r/d_{jet}=0,~x/d_{jet}=10$, likely due to an increased evaporation rate. The Adiabatic predictions in EtF4 closely align with the DCSE results, since Ref.~\cite{hussien2022simulations} employs an adiabatic Trajectory Generated Low Dimension Manifold (TGLDM) lookup table. Due to the limited amount of pre-vaporized fuel, EtF7 is more sensitive to liquid-phase properties, such as the droplet size distribution and the temperature distribution inside the droplets, rather than the multi-region flamelet model or the inlet boundary condition of $\widetilde{Z}$ at the central jet~\cite{hu2020large}. Consequently, the $\widetilde{Z}$ profiles show negligible differences among Adiabatic, Tab5D and Tab6D.

\begin{figure}[htb]
  \centering
    \includegraphics[scale=0.42]{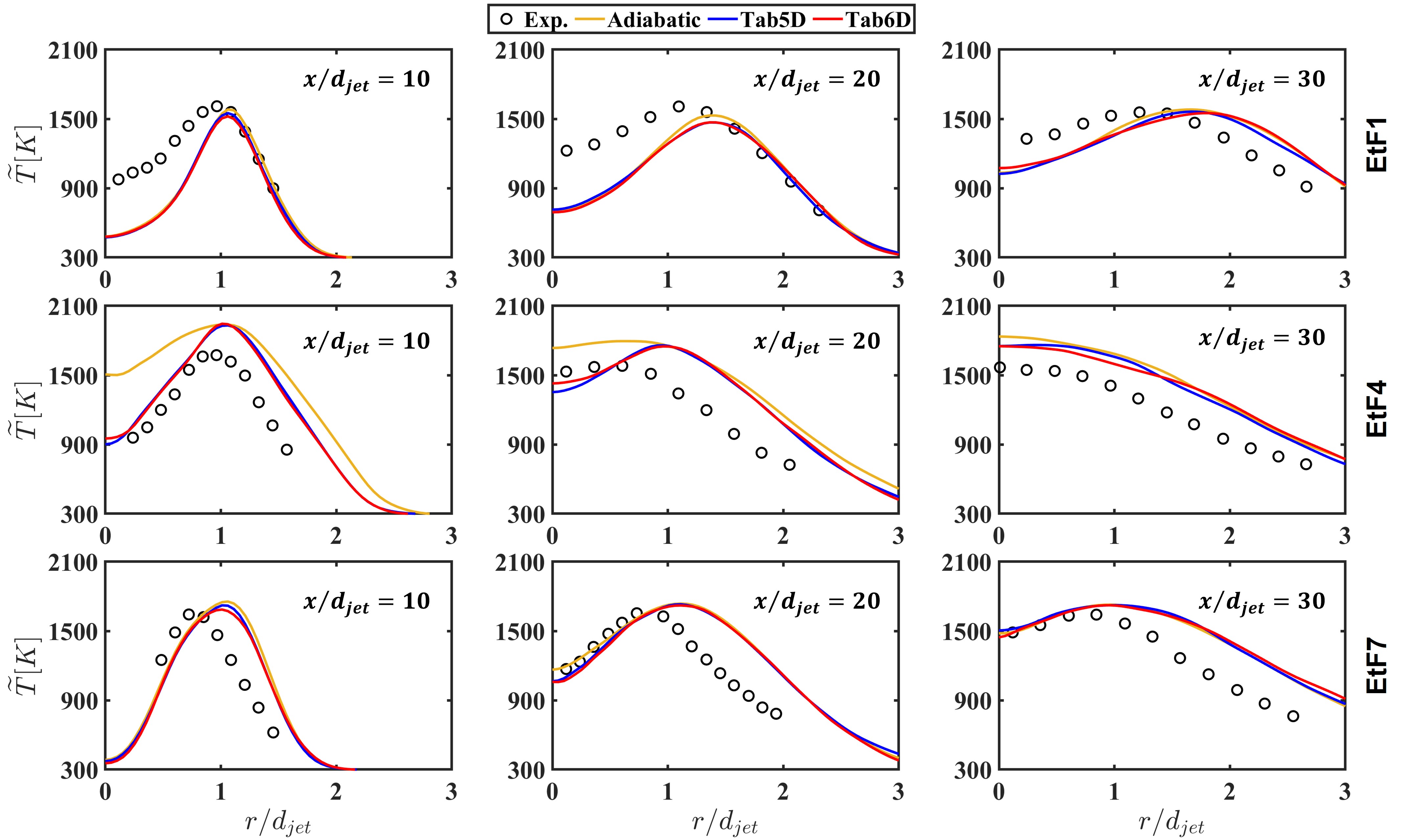}
    \caption{Radial profile of mean gas temperature $\widetilde{T}$ at different axial locations, $x/d_{jet}=10,20,30$, in the EtF1, EtF4 and EtF7 flames, compared with the experimental data~\cite{GOUNDER20123372}.}\label{fig6_temperaute}
\end{figure}

Figure~\ref{fig6_temperaute} presents the radial profile of the gas-phase temperature $\widetilde{T}$ at various axial locations for the EtF1, EtF4, and EtF7 flames, compared to the experimental data~\cite{GOUNDER20123372}. In EtF1, the flame width in the simulations agrees well with the measurements, but the temperature along the axis is lower than that observed experimentally. This phenomenon was also observed in previous simulations~\cite{kirchmann2021two, hussien2022simulations}.  This discrepancy is due to the rich gas-phase ethanol of the central jet, as discussed in Figs.~\ref{fig3_instant_T_omega}-\ref{fig5_Z}. Notably, the temperature gap between experiments and simulations narrows down from $x/d_{jet}=10$ to $x/d_{jet}=30$, suggesting that the total mass flow rate of two-phase ethanol in the central jet of EtF1 aligns with the measurements in the ultrasonic nebulizer~\cite{GOUNDER20123372}. Kirchmann et al.~\cite{kirchmann2021two} proposed that the slow temperature rise along the axis in EtF1 may be due to insufficient mixing in the pilot flow and the central jet, and thus considered an alternative setup in which the computational domain includes an additional 215mm long pipe upstream of the jet exit. However, such a setup only results in a slight temperature difference, likely because the liquid coating in the central jet was not captured~\cite{kirchmann2021two}. Another possible source of temperature discrepancies could be attributed to the limitations of thermocouple measurement techniques~\cite{rittler2015sydney, hussien2022simulations, ukai2013cmc}. Although thermocouples provide valuable time-averaged data for turbulent flames, their limited spatial resolution, reaction quenching uncertainties, and flame disturbances can introduce temperature measurement errors of up to 10\%~\cite{giusti2019turbulent, ukai2013cmc}. Furthermore, droplet collisions with thermocouples may induce premature evaporation. In EtF1, the higher droplet injection in the central jet increases the likelihood of droplet-thermocouple collisions compared to EtF4. In EtF7, the absence of thermocouples within $r/d_{jet}=5$ at $x/d_{jet}=10$ partially reduces their disturbance. Additionally, the maximum temperature at $x/d_{jet}=10$ in EtF4 is over-predicted both in the current study and in Refs.~\cite{kirchmann2021two, hussien2022simulations}. This discrepancy may result from heat transfer through the pipe wall, intensified by the low velocity and cooling effects from spray generation in the central jet.

Because the chemical reaction slows down in the presence of evaporation, Adiabatic predicts higher temperature profiles than Tab5D and Tab6D. Over-predicted temperatures accelerate liquid evaporation, increasing the mixture fraction, as shown in Fig.~\ref{fig5_Z}. In EtF1, the reactions occur at a rich mixture fraction, high progress variable and negative covariance zone at $r/d_{jet}=0.5$, $x/d_{jet}=10$, leading to a slightly higher temperature prediction by Tab5D than that of Tab6D. In EtF4, the gas mixture is in low mixture fraction, low progress variable and positive covariance along the axis. Thus, the temperature of Tab6D is slightly higher than that of Tab5D at $x/d_{jet}=10$, and then has a noticeable improvement at $x/d_{jet}=20$. At $x/d_{jet}=30$, the temperature profiles are similar since most ethanol has burnt out. Upstream of $x/d_{jet}=10$ in EtF7, combustion primarily occurs when droplets evaporate and ignite in the hot pilot flow, resulting in a high mixture fraction, high progress variable, and negative covariance. Consequently, the peak temperature using Tab6D is slightly lower than that using Tab5D and aligns with experimental results.

\subsection{Liquid phase statistics}\label{sec_liquidPhase}

\begin{figure}[htb]
  \centering
    \includegraphics[scale=0.42]{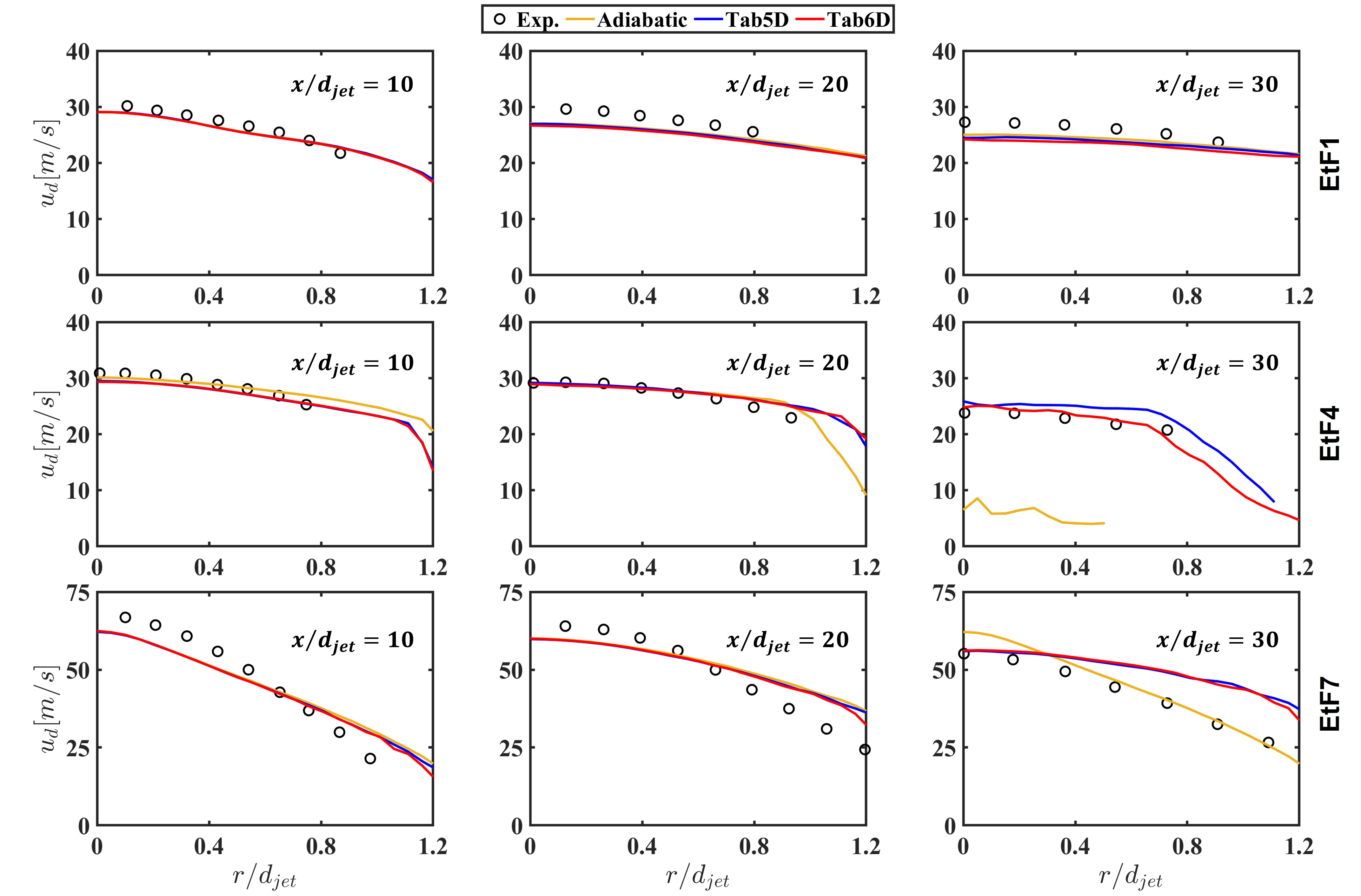}
    \caption{Radial profile of droplet axial mean velocity $u_d$ at different axial locations, $x/d_{jet}=10,20,30$, in the EtF1, EtF4 and EtF7 flames, compared with the experimental data marked as "Allsizes"~\cite{GOUNDER20123372}.}\label{fig7_dropletU}
\end{figure}

\begin{figure}[htb]
  \centering
    \includegraphics[scale=0.42]{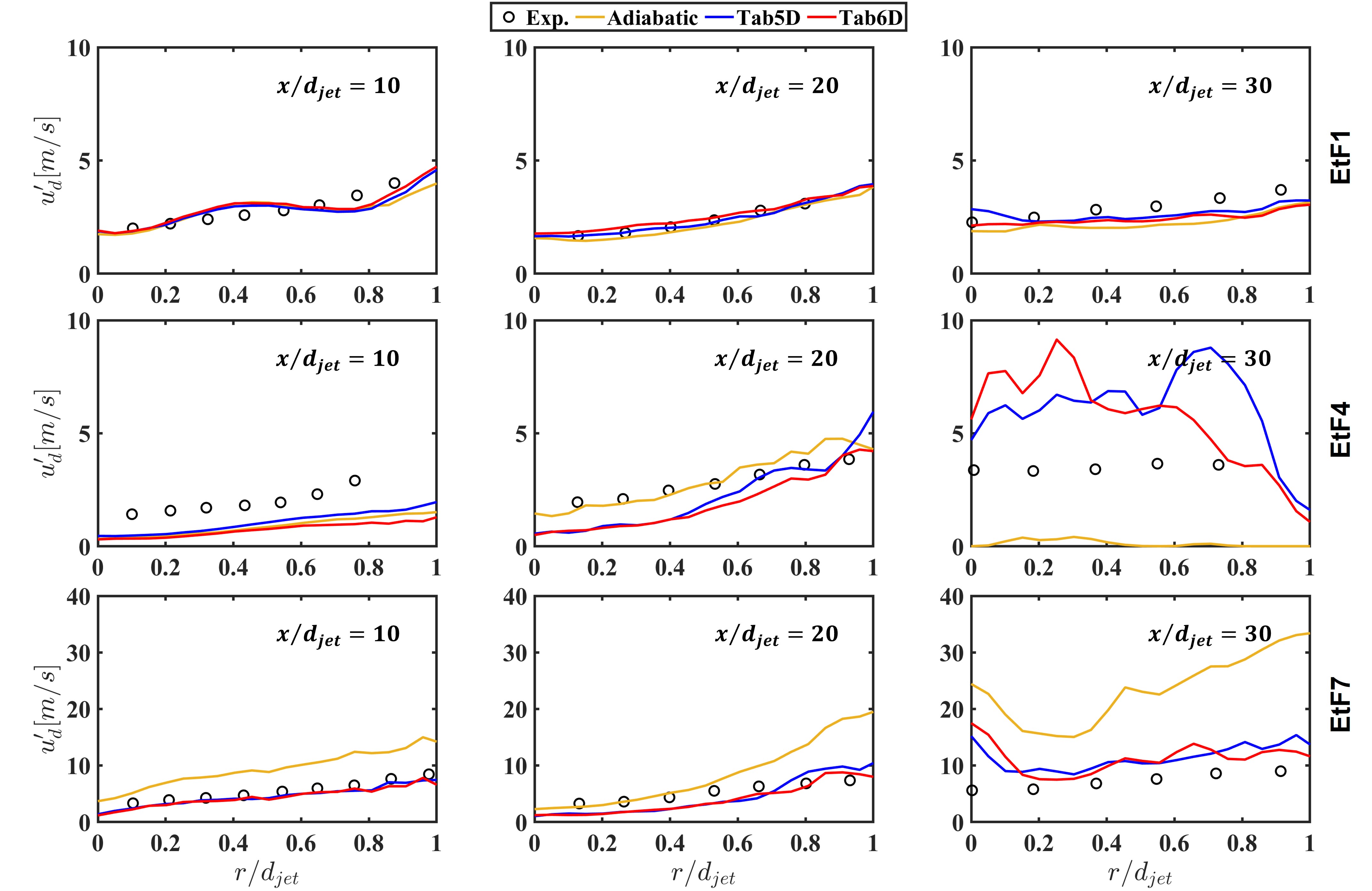}
    \caption{Radial profile of droplet RMS axial velocity $u'_d$ at different axial locations, $x/d_{jet}=10,20,30$, in the EtF1, EtF4 and EtF7 flames, compared with the experimental data marked as "Allsizes"~\cite{GOUNDER20123372}.}\label{fig8_dropletUrms}
\end{figure}

\begin{figure}[htb]
  \centering
    \includegraphics[scale=0.42]{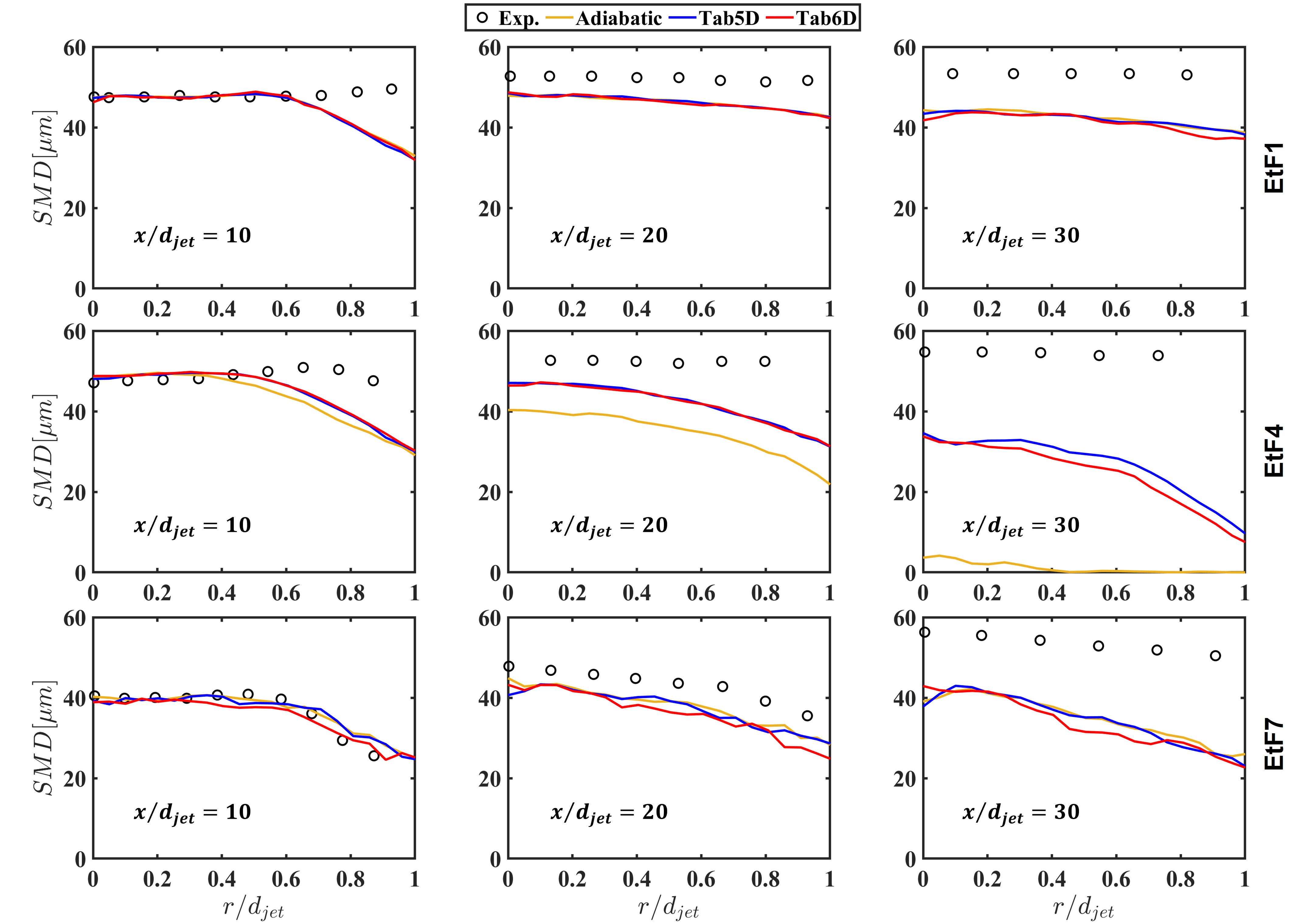}
    \caption{Radial profile of droplet SMD at different axial locations, $x/d_{jet}=10,20,30$, in the EtF1, EtF4 and EtF7 flames, compared with the experiments~\cite{GOUNDER20123372}.}\label{fig9_dropletRMS}
\end{figure}

The droplet properties, including $\mathbf{x}_d$, $m_d$, $\mathbf{u}_d$ and $T_d$, are related to the gas-phase properties such as density $\rho$, velocity $\mathbf{u}$ and temperature $T$, as described by Eqs.~(\ref{droplet_displacement})-(\ref{droplet_energy}). Fig.~\ref{fig7_dropletU} presents the radial profile of droplet axial mean velocity $u_d$ for all droplet sizes. The simulated droplet velocities align well with experimental data at the axial locations of $x/d_{jet}=10, 20, 30$. After exiting the jet exit plane, the droplets experience deceleration due to the spreading of the flame. Gas-phase combustion accelerates the gas mixture, which increases the velocity of droplets through drag forces. In EtF1, the lower temperature along the axis results in slower axial velocities. In EtF4, the high axial temperature predicted for the Adiabatic case at $x/d_{jet}=10$ accelerates the droplet axial velocity. The inclusion of covariance (Tab6D) shows slight differences in droplet velocity at $x/d_{jet}=10$ compared to Tab5D. However, improvements in gas temperature are revealed in the droplet velocity profile at $x/d_{jet}=30$. At $x/d_{jet}=10, 20$ in EtF7, the droplet velocity is under-predicted at $r/d_{jet}=0$ and over-predicted at $r/d_{jet}=1.2$, which is consistent with Fig.~\ref{fig6_temperaute}.

Fig.~\ref{fig8_dropletUrms} shows the RMS axial velocity $u'_d$ of the liquid phase. The predictions show reasonable agreement with the experimental data. In EtF4, the RMS velocity is underestimated at $x/d_{jet}=10, 20$ and overestimated at $x/d_{jet}=30$. The discrepancies between Adiabatic, Tab5D, and Tab6D in droplet RMS velocity may be attributed to differences in particle diameters, as smaller droplets are accelerated more rapidly by the gas carrier~\cite{rittler2015sydney}.

\begin{figure}[htb]
  \centering
    \includegraphics[scale=0.42]{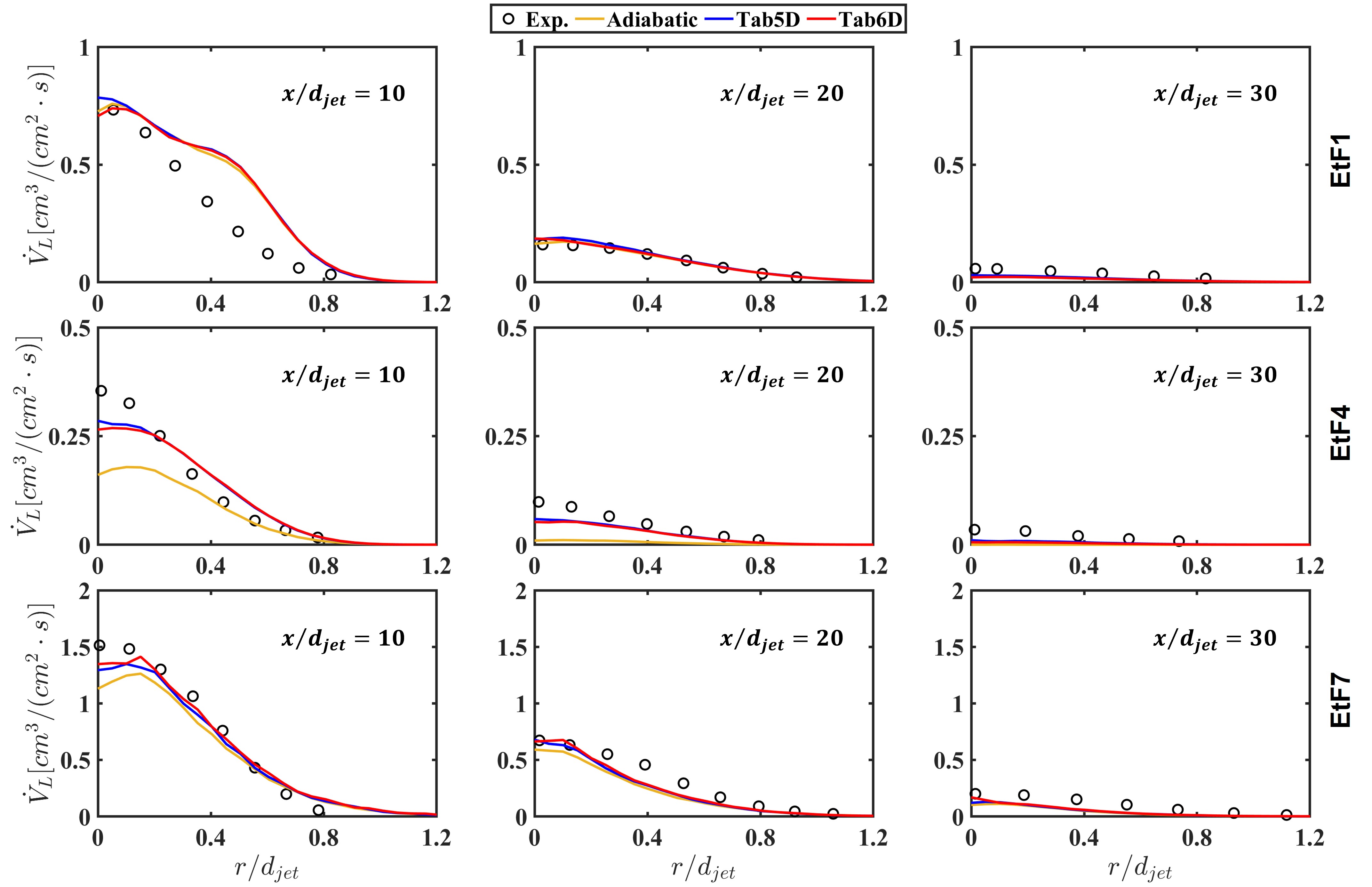}
    \caption{Radial profile of liquid volume flux at different axial locations, $x/d_{jet}=10,20,30$, in the EtF1, EtF4 and EtF7 flames, compared with the experimental data~\cite{GOUNDER20123372}.}\label{fig10_liquidVolumeFlux}
\end{figure}

The radial profiles of the Sauter mean diameter (SMD) are presented in Fig.~\ref{fig9_dropletRMS}. In EtF4, the droplet sizes of Adiabatic are significantly smaller than Tab5D and Tab6D, due to the over-prediction of central jet temperature (see Fig.~\ref{fig6_temperaute}). The SMD results align with experimental measurements at the axial location of $x/d_{jet}=10$ within $r/d_{jet}=0.6$ for all three investigated flames, but show a decrease outside $r/d_{jet}=0.6$ for EtF1 and EtF4. At $x/d_{jet}=20, 30$, the droplet diameters are under-predicted for all the three flames. Those trends have been observed in Refs.~\cite{rittler2015sydney, hussien2022simulations, kirchmann2021two} as well. Rittler et al.~\cite{rittler2015sydney} suggested that this discrepancy may be related to the faster evaporation rates of smaller particles compared to larger droplets that are surrounded by hot gas. Another potential reason is the liquid coating, as discussed in Refs.~\cite{kirchmann2021two, hussien2022simulations}. The adhered liquid film in the central jet breaks up at the nozzle, generating droplets larger than predicted by the simulations. Strong turbulence in EtF7 adequately breaks up the liquid film. Thus, the predictions on SMD align well with the measurements.

Fig.~\ref{fig10_liquidVolumeFlux} shows the radial profiles of liquid volumetric flux $\dot{V}_L$. In EtF1, the liquid volume flux is overestimated at $x/d_{jet}=10$ since the gas temperature profile is relatively low and thus reduces the evaporation rates. In EtF4, the simulated $\dot{V}_L$ is lower than the measurements near the axis, especially for Adiabatic, which is in accord with the higher temperature in Fig.~\ref{fig6_temperaute}. The predicted $\dot{V}_L$ in EtF7 aligns well with the experimental data at the axial locations investigated.

\subsection{Flame regimes and local two-phase interactions}\label{sec_flameRegime}

\begin{figure}[htb]
  \centering
    \includegraphics[scale=0.45]{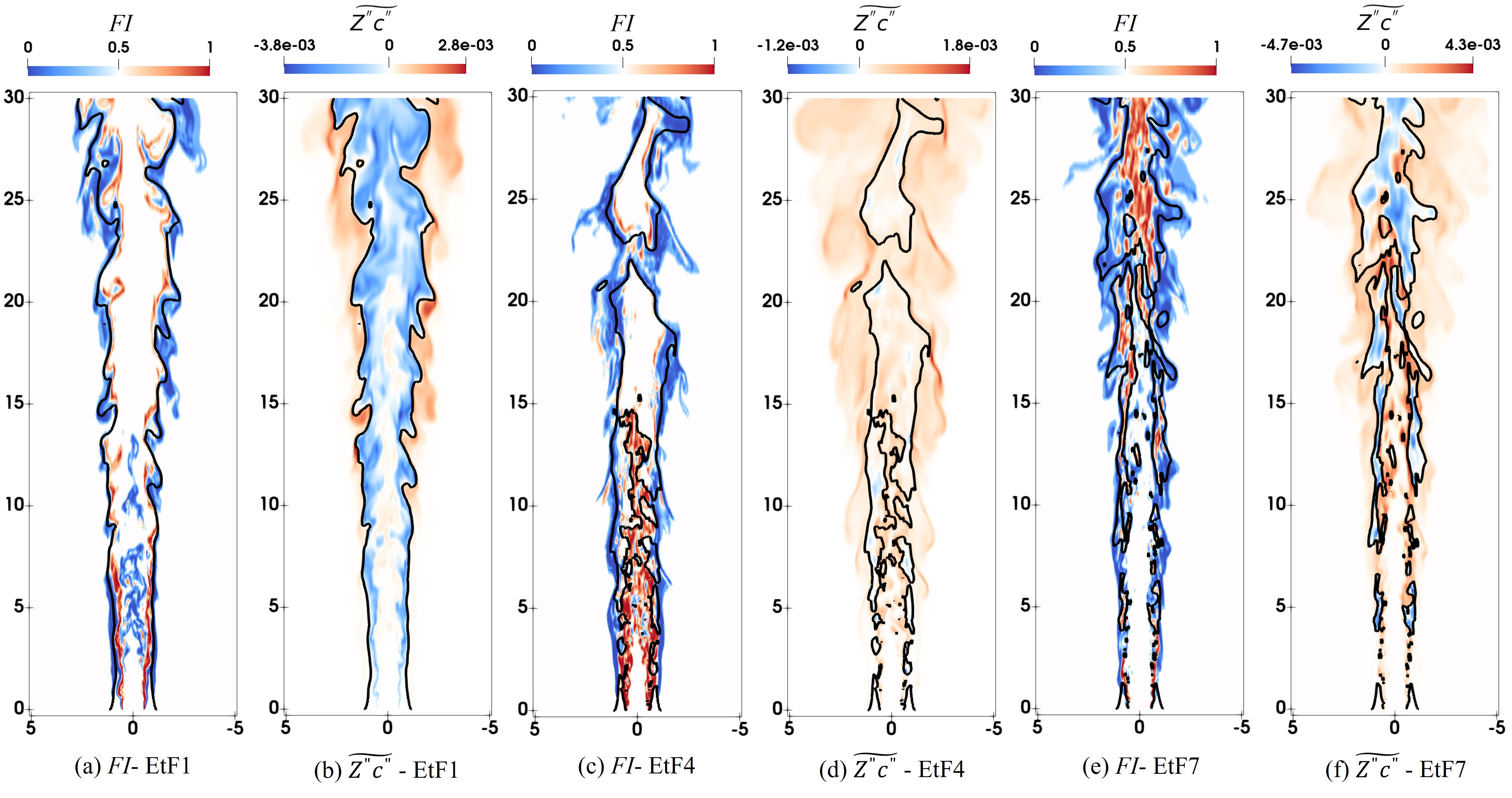}
    \caption{Instantaneous snapshots of flame index $FI$ and covariance $\widetilde{Z^{''}c^{''}}$ in the EtF1, EtF4 and EtF7 flames. The $\widetilde{Z}=\widetilde{Z}_{st}=0.1$ contour (black) is plotted at the stoichiometric mixture fraction.}\label{fig11_flameIndex}
\end{figure}

In partially premixed combustion, it is important to figure out the fraction of flamelets that are locally premixed or non-premixed regimes, which can be indicated by the flame index $FI$~\cite{rosenberg2015flame}. To identify the burning regimes in spray flames, the flame index can be defined as~\cite{hu2019partially}
\begin{equation}\label{eqn_flameIndex}
\begin{aligned}
& FI = \frac{1}{2} \left( 1 + \frac{\triangledown \overline{Y}_{Fuel} \cdot \triangledown \overline{Y}_{O2}}{|\triangledown \overline{Y}_{Fuel}| |\triangledown \overline{Y}_{O2}|} \right),
\end{aligned}
\end{equation}
where $\overline{Y}_{Fuel}$ and $\overline{Y}_{O2}$ are the mass fractions of fuel and oxygen, respectively. Thus, $FI$ is within $(0.5,1]$ for a premixed zone where the fuel and oxidizer gradients are aligned. For a non-premixed regime, $FI$ is within $[0,0.5)$ since the fuel and air are on the opposite sides of the flamelet. If no chemical reaction exists, $FI=0.5$.

Fig.~\ref{fig11_flameIndex} presents the flame index $FI$ and covariance $\widetilde{Z^{''}c^{''}}$, contoured by the stoichiometric mixture fraction $\widetilde{Z}_{st}$. The premixed and non-premixed regimes are located closely in the three flames. From the perspective of the central jet, the premixed flame front exists close to the axis, followed by a non-premixed reacting zone. The reason is that droplet evaporation produces a considerable amount of ethanol vapor, leading to diffusion-type reactions. In EtF1 and EtF7, the overall equivalent ratios in the central jet are 4.5 and 1.8 respectively, thus combustion was mainly in non-premixed mode as the ethanol is rich for the air carrier in the central jet. In EtF7, the mass flow rate of pre-vaporized fuel is only 2g/min, resulting in a thin premixed regime in Fig.~\ref{fig11_flameIndex}(e). In EtF4, the premixed regime is dominant due to the overall equivalent ratio of 1.4 in the central jet. The pre-evaporated spray jet undergoes three regimes in the covariance $\widetilde{Z^{''}c^{''}}$ downstream: ($\romannumeral1$) the mixture is ignited ($\widetilde{c}$ increases), while droplets evaporate ($\widetilde{Z}$ increases), leading to a positive $\widetilde{Z^{''}c^{''}}$ zone; ($\romannumeral2$) droplets evaporate ($\widetilde{Z}$ increases) as the mixture is almost burnt ($\widetilde{c}$ decreases), resulting in a negative $\widetilde{Z^{''}c^{''}}$ area; ($\romannumeral3$) the mixture diffuses ($\widetilde{Z}$ and $\widetilde{c}$ decrease) as ethanol in the gas and liquid phases burns out, producing a positive $\widetilde{Z^{''}c^{''}}$ regime. The effect of $\widetilde{Z^{''}c^{''}}$ is weak in EtF1 due to its thin main reaction zone in Fig.~\ref{fig3_instant_T_omega}(b). In EtF4, $\widetilde{Z^{''}c^{''}}$ has a significant impact on the temperature $\widetilde{T}$, owing to the wide positive $\widetilde{Z^{''}c^{''}}$ zone. In EtF7, reactions mainly occur in the negative $\widetilde{Z^{''}c^{''}}$ regimes, as shown in Fig.~\ref{fig3_instant_T_omega}. Thus, the distribution of $\widetilde{T}$ is well predicted using Tab6D.

\begin{figure}[htb]
  \centering
    \includegraphics[scale=0.11]{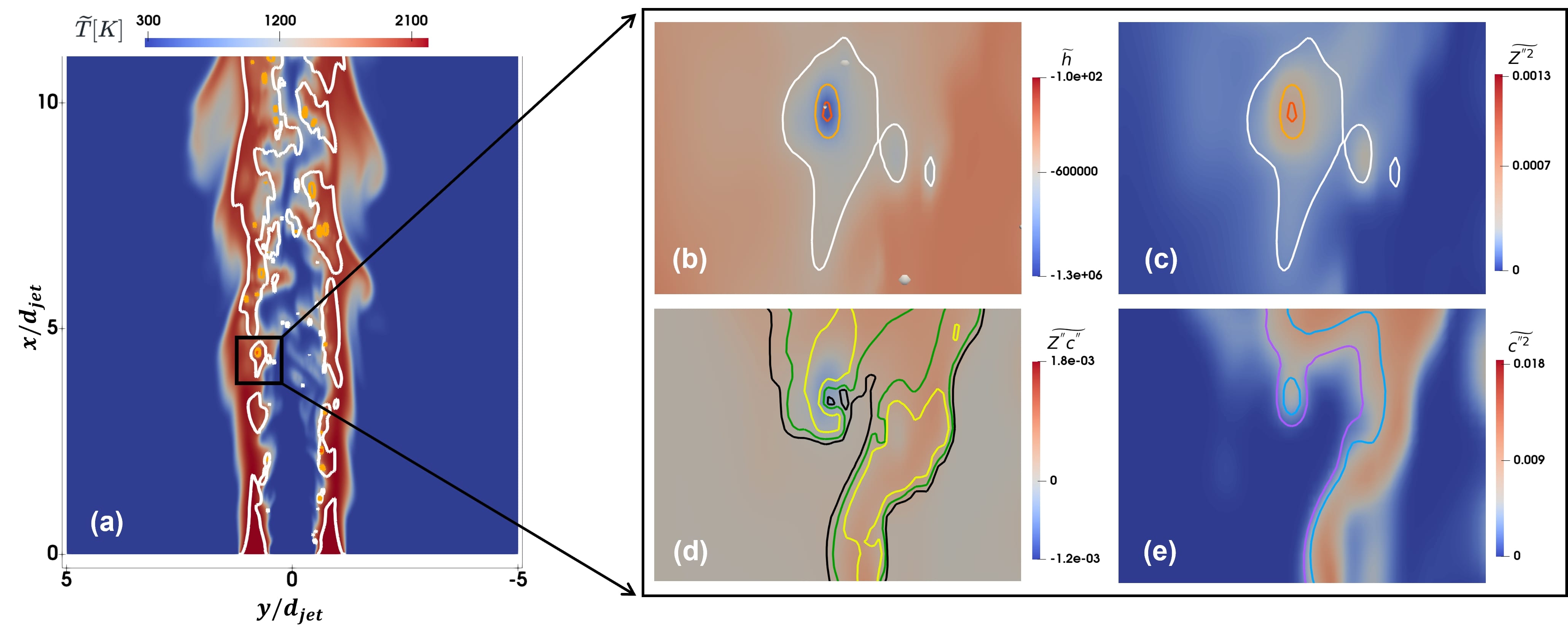}
    \caption{Instantaneous snapshots of (a) $\widetilde{T}$, (b) $\widetilde{h}$, (c) $\widetilde{Z^{''2}}$, (d) $\widetilde{Z^{''} c^{''}}$ and (d) $\widetilde{c^{''2}}$ in EtF4. Contours in (a)-(c) are plotted as: $\widetilde{Z}=\widetilde{Z}_{st}=0.1$ (white), $\widetilde{Z}=0.15$ and $\widetilde{Z}=0.2$. Contours in (d) are plotted as:  $\widetilde{\dot{\omega}}_c=10$ (black), $\widetilde{\dot{\omega}}_c=20$ (green) and $\widetilde{\dot{\omega}}_c=40$ (yellow). Contours in (e) are plotted as: $\widetilde{c}_n=0.8$ (blue) and $\widetilde{c}_n=0.9$ (purple). }\label{fig12_localEtf4}
\end{figure}

To illustrate the interaction among droplet evaporation, heat loss and covariance, Fig.~\ref{fig12_localEtf4} presents the local instantaneous fields in EtF4 of (a) $\widetilde{T}$, (b) $\widetilde{h}$, (c) $\widetilde{Z^{''2}}$, (d) $\widetilde{Z^{''} c^{''}}$ and (e) $\widetilde{c^{''2}}$. The location of Figs.~\ref{fig12_localEtf4}(b)-(e) is highlighted with a black box in Fig.~\ref{fig12_localEtf4}(a). In Fig.~\ref{fig12_localEtf4}(b), a liquid parcel (gray sphere) evaporates, which reduces the local $\widetilde{h}$ and increases the $\widetilde{Z}$, as shown in Fig.~\ref{fig12_localEtf4}(b)). Fig.~\ref{fig12_localEtf4}(c) presents that the local increase of $\partial \widetilde{Z}/\partial x_j$ induces a high $\widetilde{Z^{''2}}$, as indicated by Eq.~(\ref{zvar_eqn}). As $\widetilde{Z}$ moves away from $\widetilde{Z}_{st}$, the reactions tend to slow down. Fig.~\ref{fig12_localEtf4}(e) illustrates that the local $\widetilde{c}$ is diluted by the vapor fuel. Fig.~\ref{fig12_localEtf4}(d) shows that the local increase of $\widetilde{Z}$ and decrease of $\widetilde{c}$ tend to reduce the covariance $\widetilde{Z^{''} c^{''}}$, as implied in Eq.~(\ref{zcvar_eqn}). Fig.~\ref{fig12_localEtf4}(e) presents that the decrease of $\widetilde{c}$ raises its variance, as suggested by Eq.~(\ref{cvar_eqn}). Influenced by $\widetilde{h}$, $\widetilde{Z}$, $\widetilde{Z^{''2}}$, $\widetilde{c}$, $\widetilde{c^{''2}}$ and $\widetilde{Z^{''} c^{''}}$, the chemical reactions are slowed down, as shown in Fig.~\ref{fig12_localEtf4}(d). Via the evaporation rate model, i.e., Eq.~(\ref{droplet_mass}), the evaporation process decelerates as $\widetilde{T}$ decreases and $\widetilde{Y}_{C2H5OH}$ increases. Moreover, rapid evaporation may induce local extinction since the flammability limits are narrowed down and the local mixture fraction increases rapidly.

\subsection{Influences of evaporation, mixing, and combustion}\label{sec_scalarInfluences}

\begin{figure}[htb]
  \centering
    \includegraphics[scale=0.45]{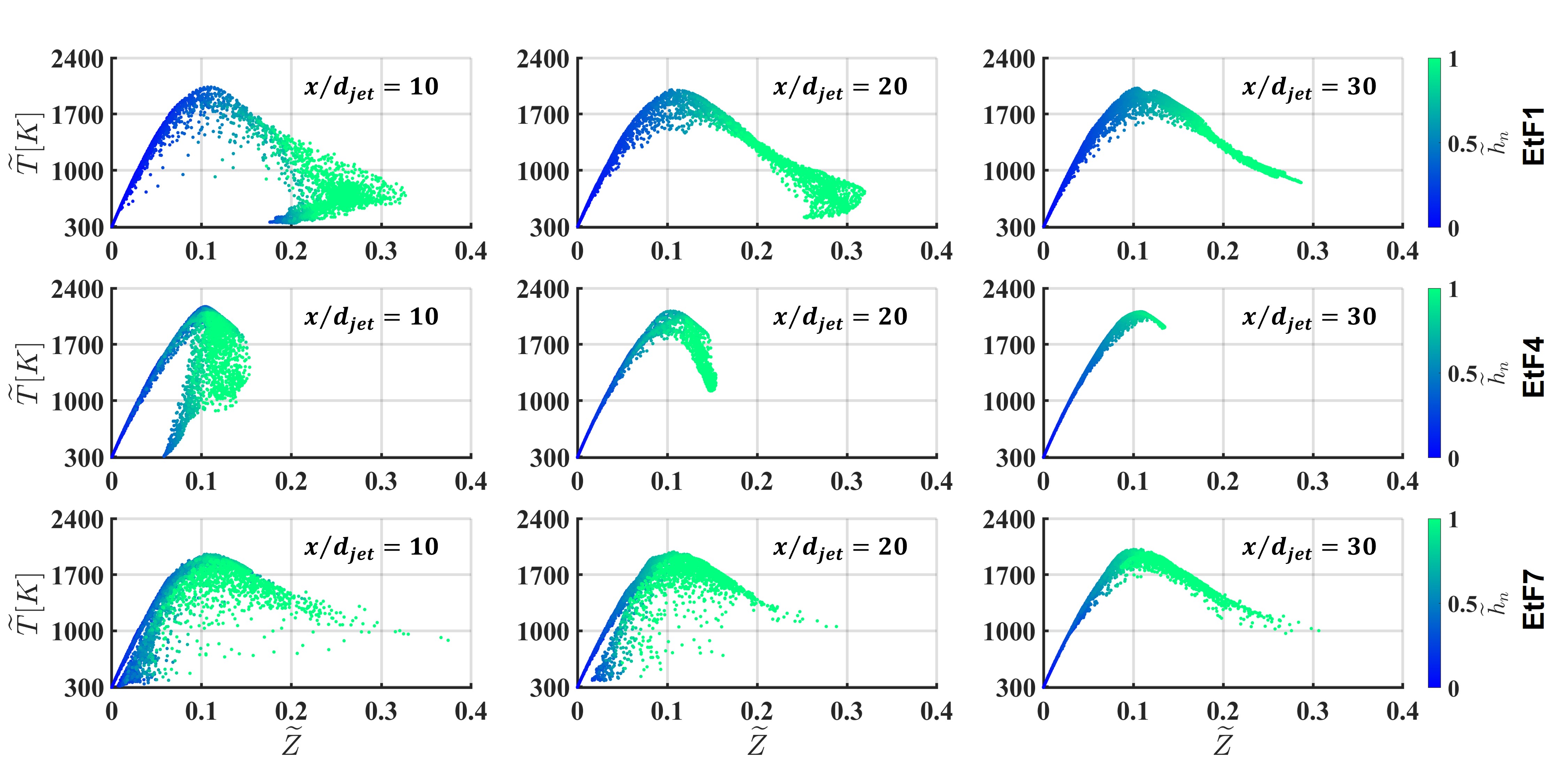}
    \caption{Scatter plots of the mean gas temperature $\widetilde{T}$ versus mixture fraction $\widetilde{Z}$ at three cross-sections, $x/d_{jet}=10,20,30$, in the EtF1, EtF4 and EtF7 flames. Points are colored by the normalized heat loss $\widetilde{h}_{n}$.}\label{fig13_scatterTZ}
\end{figure}

\begin{figure}[htb]
  \centering
    \includegraphics[scale=0.45]{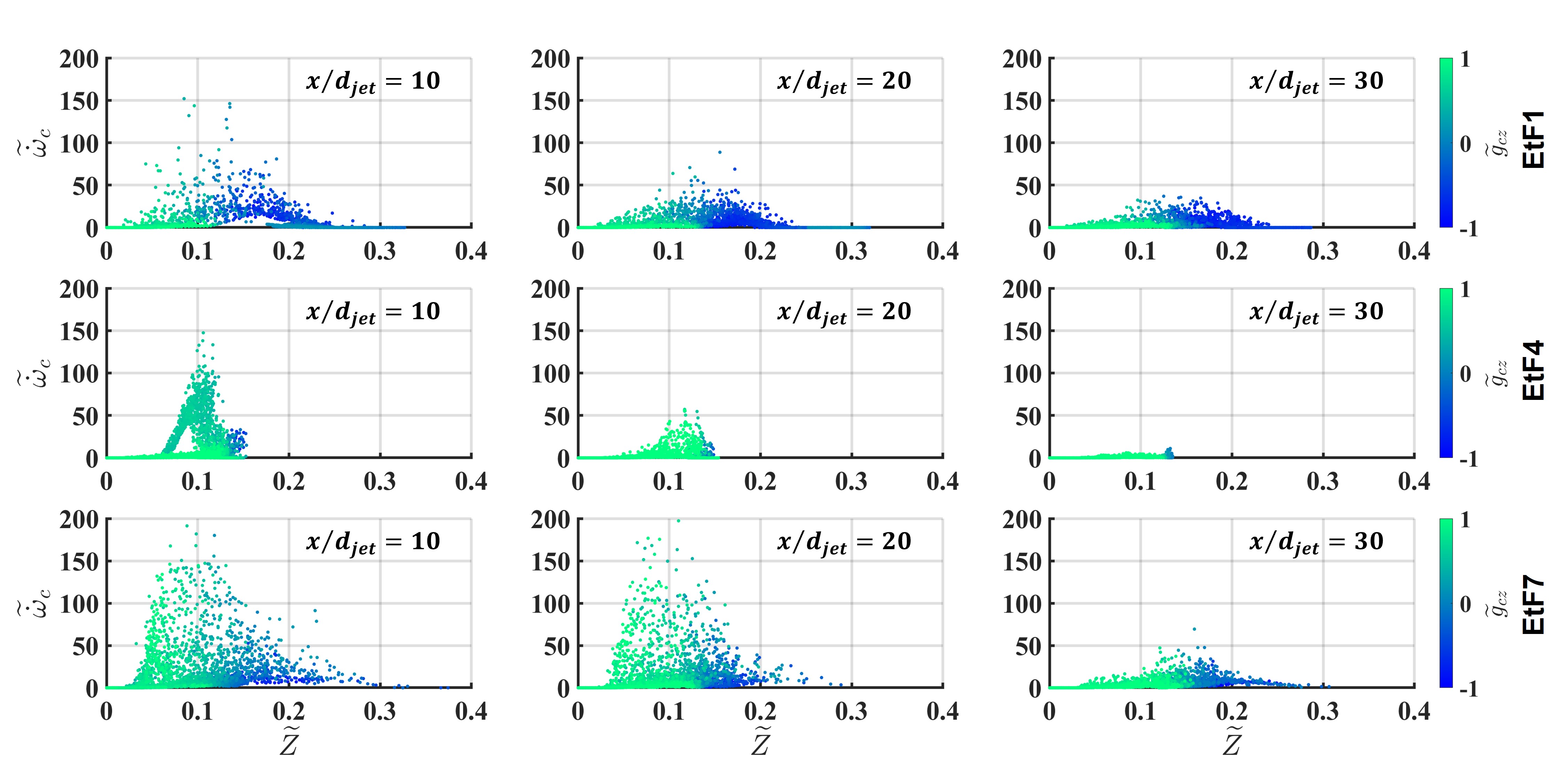}
    \caption{Scatter plots of the chemical reaction rate $\widetilde{\dot{\omega}}_c$ versus mixture fraction $\widetilde{Z}$ at three cross-sections, $x/d_{jet}=10,20,30$, in the EtF1, EtF4 and EtF7 flames. Points are colored by the scalar correlation $\widetilde{g}_{cz}$.}\label{fig14_scatteromegacZ}
\end{figure}

The mixing and thermochemical states of the gas mixture at different axial locations can be analyzed using scatter plots of $\widetilde{T}$ versus $\widetilde{Z}$~\cite{meier2006investigations}, as presented in Fig.~\ref{fig13_scatterTZ}. Data points are colored by the normalized evaporation-induced specific enthalpy reduction $\widetilde{h}_{n} = \widetilde{h}_{r} / \widetilde{h}_{r, max}$, where $\widetilde{h}_{r, max} = 9.9\times10^{4}$J/kg. The large amount of pre-vaporized fuel in EtF1 ($29.3$ g/min) leads to slower chemical reactions, resulting in some gas mixture characterized by low temperature, high mixture fraction, and low enthalpy reduction at $x/d_{jet} = 10$. Further downstream, the pre-vaporized ethanol is gradually heated and participates in combustion. In EtF4, combustion primarily occurs near $\widetilde{Z}_{st}$, leading to faster chemical reactions and higher peak temperatures compared to EtF1. In EtF7, the $\widetilde{Z}$ distribution may result from the strong turbulence (jet Reynolds number $Re = 45,700$). At $x/d_{jet} = 30$, the $\widetilde{T}$–$\widetilde{Z}$ distribution resembles that of EtF1, as these two flames share the same total fuel mass flow rate ($75$ g/min).

Figure~\ref{fig14_scatteromegacZ} depicts the distribution of the chemical reaction rate $\widetilde{\dot{\omega}}_c$ with respect to the mixture fraction $\widetilde{Z}$, colored by the scalar correlation $\widetilde{g}_{cz}$. Negative values of $\widetilde{g}_{cz}$ occur at higher mixture fractions, which can be attributed to droplet evaporation. At $x/d_{jet}=10$, $\widetilde{\dot{\omega}}_c$ reaches its peak at $\widetilde{Z}_{st}=0.1$. However, at $x/d_{jet}=20,30$, the maximum $\widetilde{\dot{\omega}}_c$ shifts toward higher $\widetilde{Z}$ than $\widetilde{Z}_{st}$. This trend may be explained by the contribution of scalar correlation, as strong correlation can also increase the peak $\widetilde{\dot{\omega}}_c$ (see Fig.\ref{fig2_omegac_tab}). Compared with EtF1 and EtF7, the EtF4 flame exhibits faster reaction rates that concentrate around $\widetilde{Z}_{st}$, coupled with a positive $\widetilde{g}_{cz}$. Consequently, EtF4 demonstrates heightened sensitivity to scalar correlation in terms of gas temperature.

\begin{figure}[htb]
  \centering
    \includegraphics[scale=0.5]{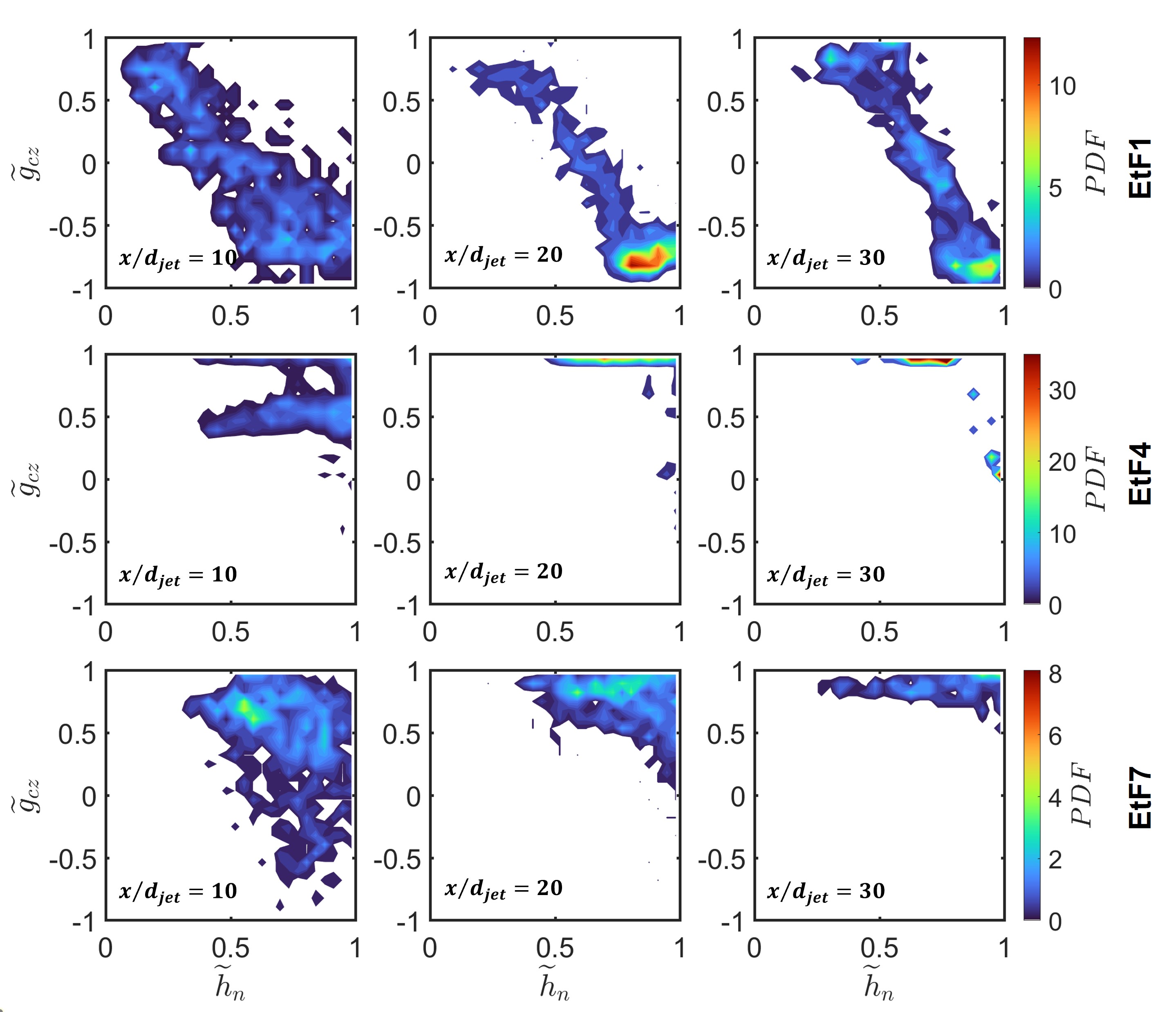}
    \caption{The joint PDF of the normalized evaporation-induced specific enthalpy reduction $\widetilde{h}_{n}$ and the scalar correlation $\widetilde{g}_{cz}$ at three cross-sections, $x/d_{jet}=10,20,30$, in the EtF1, EtF4 and EtF7 flames. Points are colored by $\widetilde{g}_{cz}$.}\label{fig15_pdfHngcz}
\end{figure}

Fig.~\ref{fig15_pdfHngcz} presents the instantaneous joint PDF of $\widetilde{h}_{n}$ and $\widetilde{g}_{cz}$ in the EtF1, EtF4 and EtF7 flames. The data are collected across the entire plane at the three axial locations from the Tab6D simulation. To investigate the effects of evaporation, reaction kinetics and turbulent mixing, the dataset is preprocessed by filtering out low-reactivity regions ($\widetilde{\dot{\omega}}_c < 3$). In these spray flames, $\widetilde{h}_{n}$ increases as fuel droplets evaporate, while $\widetilde{g}_{cz}$ increases if $\widetilde{Z}$ and $\widetilde{c}$ rise or fall simultaneously. The evaporation of liquid ethanol droplets produces two effects: ($\romannumeral1$) it increases the mass fraction of ethanol, thereby raising $\widetilde{Z}$ substantially; and ($\romannumeral2$) it dilutes the existing gas mixture with newly evaporated ethanol, lowering the mass fractions of combustion products, thus reducing $\widetilde{c}$. Together, these two effects drive the correlation coefficient $\widetilde{g}_{cz}$ to negative values, which predominantly occur in regions characterized by a high $\widetilde{h}_{n}$. However, the chemical reaction enhances $\widetilde{c}$, thus increasing $\widetilde{g}_{cz}$ as $\widetilde{Z}$ increases. As a result, in the presence of droplet evaporation, the value of $\widetilde{g}_{cz}$ is determined by the competing effects of phase change and combustion on $\widetilde{c}$. In EtF1, because the gas mixture is excessively fuel-rich, the chemical reaction proceeds slowly, as shown in Fig.~\ref{fig14_scatteromegacZ}. Consequently, evaporation dominates the $\widetilde{g}_{cz}$ field, leading to strong negative correlations between $\widetilde{h}_{n}$ and $\widetilde{g}_{cz}$, with Pearson correlation coefficients of -0.8, -0.84, and -0.85 at $x/d_{jet} = 10, 20, 30$, respectively. In contrast, in EtF4, the intensified chemical reaction results in a wide range of positive $\widetilde{g}_{cz}$ distribution, as shown in Fig.~\ref{fig11_flameIndex}(d). Turbulent mixing influences combustion and evaporation through multi-scale vortical interactions. It accelerates fuel-oxidizer homogenization by breaking down concentration gradients. Simultaneously, turbulence induces flame wrinkling, amplifying the reactive interface area to enhance global reaction rates. However, excessive strain rates from intense eddies may surpass the flame’s extinction limit, triggering localized quenching, as in the EtF7 flame shown in Fig.~\ref{fig2_omegac_tab}(f). Moreover, at the jet exit plane in EtF7, 97\% of the injected fuel remains in the liquid phase, contributing to the negative values of $\widetilde{g}_{cz}$. Meanwhile, intense chemical reactions promote an increase in $\widetilde{c}$, leading to positive $\widetilde{g}_{cz}$ in regions with high $\widetilde{h}_{n}$.

\section{Conclusions}\label{sec_conclusions}

This study proposed a modeling approach of turbulent spray combustion by integrating the high-dimensional FGM method to evaluate the effects of evaporation-induced specific enthalpy reduction $\widetilde{h}_{r}$ and scalar correlation $\widetilde{g}_{cz}$. A novel joint presumed PDF method grounded in copula theory was developed and used, in order to capture SGS correlations between mixture fraction $\widetilde{Z}$ and progress variable $\widetilde{c}$. Rapid grid convergence was achieved in the joint PDF method. 

Effects of $\widetilde{h}_{r}$ and $\widetilde{g}_{cz}$ on chemical reaction rate $\widetilde{\dot{\omega}}_c$ were analyzed. As $\widetilde{h}_{r}$ increases due to evaporation, the chemical reactions decelerate, and the range of flammability limits is narrowed down. The impact of $\widetilde{g}_{cz}$ on the $\widetilde{\dot{\omega}}_c$ was two-fold: ($\romannumeral1$) increasing the peak value if $\widetilde{g}_{cz} \neq 0$, and ($\romannumeral2$) changing the slope in the space of $\widetilde{Z}$ and $\widetilde{c}$.

The Sydney ethanol spray flames (EtF1, EtF4 and EtF7) were numerically investigated, showing good agreement with experimental data. Scalar dissipation models were employed in the multiphase reacting flows. The following key insights were drawn:
\begin{itemize}
\item The modeling of $\widetilde{h}_{r}$ and $\widetilde{g}_{cz}$ improved the gas-phase predictions, especially in spray flames characterized by intense premixed combustion of the stoichiometric mixture. The modeling of $\widetilde{h}_{r}$ demonstrated superior improvements in predicting gas temperatures compared to the modeling of $\widetilde{g}_{cz}$. The reason was that the adiabatic assumption overestimated chemical reaction rates. In contrast, the impact of scalar correlation on chemical reaction, whether accelerating or decelerating, was critically governed by the local thermochemical state, particularly the interplay between $\widetilde{Z}$ and $\widetilde{c}$.
\item The properties of liquid-phase, i.e., axial mean/RMS velocity, SMD and volume flux, were found to be modulated by gas-phase temperature, velocity and density. Due to the indirect relation, the modeling of $\widetilde{g}_{cz}$ showed minimal improvements in the liquid-phase prediction.
\item Under identical air carrier mass flow rates, elevating the liquid fuel loading (as exemplified by EtF1 versus EtF4) was found to increase the spatial dominance of the non-premixed combustion regime, with the covariance $\widetilde{Z^{''} c^{''}}$ exhibiting increased negative values due to enhanced droplet evaporation effects. Turbulent mixing enhanced energy and species transport, but excessively intense eddies may trigger local quenching, as observed in EtF7.
\item In the three flames, local evaporation and combustion exhibited competing effects on $\widetilde{g}_{cz}$. Droplet evaporation introduced fresh fuel to gas phase, increasing $\widetilde{Z}$ and decreasing $\widetilde{c}$. On the contrary, chemical reactions elevated the progress variable. Consequently, strong negative correlations between $\widetilde{g}_{cz}$ and $\widetilde{h}_{r}$ were found in EtF1, since combustion was suppressed by the excessively rich fuel.
\item The central jet temperature profiles in EtF1 exhibited systematic underprediction in both the present study and prior simulations. Analysis of $\widetilde{Z}$ and OH distributions revealed that this thermal discrepancy was probably attributed to ethanol vapor maldistribution originating from wall-adhered droplets near the jet nozzle. These findings prioritize droplet-wall interaction diagnostics in future spray combustion experiments.
\end{itemize}

In conclusion, this work systematically examined various subgrid dynamics of turbulent mixing, combustion, evaporation, specific enthalpy reduction and scalar correlation. Note that copula functions, which characterize the $\widetilde{Z}$-$\widetilde{c}$ correlations, might vary depending on fuel Lewis numbers, flow configurations (jet/swirl/shear layer-driven/strained flow), and combustion modes (premixed/diffusion/ partially premixed). This will be the focus of future investigations.

\section*{Declaration of competing interest} \label{declareOfIntrest}

The authors declare that they have no known competing financial interests or personal relationships that could have appeared to influence the work reported in this paper.

\section*{Acknowledgements}
\label{Acknowledgments}
This work is supported by the National Natural Science Foundation of China (Grant Nos. 92270203 and 52276096). Part of the numerical simulations was performed on the High-Performance Computing Platform of CAPT of Peking University.

\bibliography{cas-refs}

\end{document}